\documentclass{aa}  

\usepackage{graphicx}
\usepackage{lipsum}
\usepackage{tikz}
\usetikzlibrary{arrows.meta, positioning}
\usepackage{subcaption}         
\usepackage{lscape}             
\usepackage{placeins}           
\usepackage{hyperref}[breaklinks=true, colorlinks=true]
\usepackage{tabularx}

\usepackage{amsmath}    
\usepackage{amssymb}    

\usepackage{gensymb}
\usepackage{newtxtext,newtxmath}
\usepackage{longtable}
\usepackage{setspace}
\usepackage{booktabs}
\usepackage{wallpaper}
\usepackage{epstopdf}
\usepackage{eso-pic}
\usepackage{eqparbox}
\usepackage{multicol}
\usepackage{dcolumn}
\usepackage{listings}
\usepackage{multirow}

\usepackage[singlelinecheck=off]{caption}

\usepackage{ae,aecompl}

\newcommand{\asec}{$^{\prime\prime}$}
\def\3n{HC$_{3}$N}  
\def\5n{HC$_{5}$N}
\def\cratio{$^{12}$C/$^{13}$C}

\begin{document}

   \title{High \cratio\;isotopic ratios toward G+0.693-0.027: \\ evidence for gas inflow to the Central Molecular Zone}

 \author{L. Colzi\inst{1}
         \and
          O. Sipilä\inst{2}
          \and
          I. Jiménez-Serra\inst{1}
          \and
          V. M. Rivilla\inst{1}
          \and
          M. Sanz-Novo\inst{2}
          \and
          S. Martín\inst{3,4}
          \and
          S. Zeng\inst{5}
         }

   \institute{Centro de Astrobiología (CAB), CSIC-INTA, Ctra. de Ajalvir Km. 4, 28850, Torrejón de Ardoz, Madrid, Spain \\ \email{lcolzi@cab.inta-csic.es}
              \and
              Max-Planck-Institut für Extraterrestrische Physik (MPE), Giessenbachstr. 1, 85748 Garching, Germany
             \and
              European Southern Observatory, Alonso de Córdova 3107, Vitacura 763 0355, Santiago, Chile
              \and
              Joint ALMA Observatory, Alonso de Córdova 3107, Vitacura 763 0355, Santiago, Chile
             \and
             Star and Planet Formation Laboratory, Cluster for Pioneering Research, RIKEN, 2-1 Hirosawa, Wako, Saitama, 351-0198, Japan
             }

   \date{Received 30 April 2026; Accepted 07 July 2026}

 
  \abstract
   {Isotopic ratios are essential tools for understanding Galactic chemical evolution and reconstructing the star formation history of the Milky Way, as different isotopes are produced through distinct stellar nucleosynthesis processes. While the $^{12}$C/$^{13}$C ratio is known to increase with galactocentric distance in the Galactic disc, measurements in the Central Molecular Zone (CMZ) have historically yielded low values ($\sim$3–30) that are often hindered by high line optical depth effects.}
   {This work aims to determine the initial $^{12}\mathrm{C}/^{13}\mathrm{C}$ isotopic ratio for the parent material of the CMZ molecular cloud G+0.693$-$0.027. By employing optically thin molecular tracers and applying corrections for isotopic fractionation effects, we aim to provide a more accurate constraint on the nucleosynthetic enrichment history of the Galactic Center.} 
   {We used an ultra-high-sensitivity spectral survey carried out with the IRAM 30m and Yebes 40m telescopes to detect single and double $^{13}$C isotopologues of HC$_3$N and HC$_5$N. We analyzed the data and derived the column densities and isotopic ratios, which were then compared with astrochemical models including $^{13}$C- and $^{15}$N-isotopologues to account for potential degree of isotopic fractionation.}
   {We derived $^{12}$C/$^{13}$C ratios of 36.7$\pm$1.02 for the main velocity component of HC$_3$N (from double isotopologues) and 38.8$\pm$1.5 for HC$_5$N. These values are significantly higher than typical CMZ measurements from simpler species. Our models suggest low to intermediate fractionation degree at early timescales ($<$3$\times 10^4$ yr), leading to an inferred initial $^{12}$C/$^{13}$C ratio of $\sim$48 for the material from which present-day CMZ molecular clouds formed.}
   {The derived ratios (37--48) are consistent with those found at Galactocentric distances of 3--5 kpc. This supports a scenario where the CMZ is replenished by gas inflows from the Galactic disc, likely driven by the Galactic bar, with a possible contribution from the accretion of less-processed material from external sources such as dwarf galaxies.}


 \authorrunning{Colzi et al.}
\maketitle
\nolinenumbers

\section{Introduction}

Understanding the chemical evolution driven by stellar nucleosynthesis is essential to reconstruct the star formation history of galaxies (e.g., \citealt{maiolino2019,arcones2023}). Galactic Chemical Evolution (GCE) models are widely employed to constrain stellar nucleosynthesis processes and the formation timescales of the different Galactic components (halo, bulge, thick and thin discs; e.g., \citealt{matteucci2021}). They also provide valuable insights into the initial mass function of external galaxies (e.g., \citealt{romano2017,zhang2018}).

In the Milky Way, GCE models adopting different nucleosynthesis prescriptions predict trends that are broadly consistent with the elemental abundances measured in stellar atmospheres as a function of metallicity for stars in the solar neighborhood. However, the abundances of the less common isotopes (such as $^{13}$C, $^{15}$N, $^{18}$O, and $^{34}$S) are more difficult to determine in stars, as their spectral features are weaker than those of the main isotopes. These isotopic ratios can generally be measured only in highly evolved giant stars, where strong spectral features are detectable but the original isotopic composition has already been altered and is not pristine anymore (e.g., \citealt{tautvaisiene2016}). Measurements of $^{12}$C/$^{13}$C and $^{16}$O/$^{18}$O ratios in non-evolved stars are scarce, with only a few studies available (e.g., \citealt{botelho2020}). 
Therefore, to overcome this limitation, isotopic ratios measured toward Galactic molecular clouds are commonly employed to constrain the Galactic chemical evolution as a function of the galactocentric distance ($R_{\rm GC}$; see, e.g., \citealt{wilson1992,wilson1994,milam2005,colzi18b,yan2019,jacob2020,kobayashi2020,colzi2022b,bohm2026}).

In this work, we focus on the carbon isotopic ratio, \cratio. 
$^{12}$C is a primary element, formed directly from H and He during He-burning in the interiors of red giants and supergiants 
(e.g., \citealt{burbidge1957,woosley2002,karakas2014}). 
Conversely, $^{13}$C has a distinct origin, with a primary production channel in massive, low-metallicity rotating stars (e.g., \citealt{chiappini2008,prantzos2018}), and a secondary production pathway--requiring pre-existing $^{12}$C and $^{16}$O seeds--through the cold and hot CNO cycles operating in main-sequence stars and nova explosions, respectively (\citealt{romano2022}).
Observational studies of various molecular species, such as CN, CH, CS, CO, and H$_2$CO, toward molecular clouds located at different galactocentric distances ($R_{\rm GC}$) have revealed an increasing trend of \cratio\;with $R_{\rm GC}$, ranging from $\sim$30 at 2~kpc to $\sim$80 at 11~kpc (e.g., \citealt{milam2005,yan2019,jacob2020,yan2023}). 
These observations have been compared with GCE models by \citet{romano2017,romano2019} and \citet{colzi2022b}, which constrained stellar nucleosynthesis prescriptions in the range $R_{\rm GC}$ = 4--11~kpc. 
Their models successfully reproduce the observed linear increase of \cratio\;with $R_{\rm GC}$.

In the Central Molecular Zone (CMZ; the inner $\sim$300~pc of our Galaxy), the \cratio\;ratio has been extensively measured over the past five decades with the advent of radioastronomical observations. 
Measurements toward molecular clouds such as Sgr~B2 and Sgr~A, obtained from both absorption and emission lines, have yielded values ranging from $\sim$3 to 30 for simple molecules including HC$_3$N, H$_2$CO, OCS, HCO$^+$, HCN, HNC, C$_3$, and CH 
(\citealt{fomalont1973,wannier1978,gardner1979,goldsmith1981,gardner1982,guesten1985b,turner1991,riquelme2010,giesen2020,jacob2020}). 
Furthermore, values of $\sim$20--32 have been derived from complex organic molecules (COMs; carbon-bearing species with six or more atoms, \citealt{herbst2009}), such as methanol (CH$_3$OH), methyl cyanide (CH$_3$CN), ethanol (CH$_3$CH$_2$OH), vinyl cyanide (C$_2$H$_3$CN), and ethyl cyanide (C$_2$H$_5$CN) (e.g., \citealt{muller2008,belloche2013,belloche2016,muller2016,margules2016,halfen2017}).

These values are broadly consistent with the increase of the \cratio\;ratio with $R_{\rm GC}$ and with an extrapolation of the observed Galactic trend toward the CMZ, yielding values of $\sim$10--20. From the perspective of stellar nucleosynthesis, this behavior is expected: the higher metallicity in the CMZ could lead to a higher relative production of $^{13}$C with respect to $^{12}$C. In this scenario, $^{12}$C is primarily produced by first-generation massive stars on short timescales, whereas $^{13}$C is mainly synthesized through CNO processing of $^{12}$C in lower-mass stars over longer evolutionary timescales (e.g., \citealt{prantzos1996}).
However, \citet{wilson1994} already pointed out that the chemical evolution of the Galactic disc and that of the Galactic center may differ, and there is no basis for assuming that the abundance or isotopic gradients observed in the disc continue smoothly into the central regions. 
From an observational point of view, most of the reported \cratio\;values have been derived using the direct method, that is, by comparing the line intensities of the main isotopologue (e.g., HCO$^{+}$ or HCN) with those of its counterpart with $^{13}$C (e.g., H$^{13}$CO$^{+}$ or H$^{13}$CN). However, these measurements are strongly affected by line optical depth effects, and thus generally represent lower limits. It is worth noting that values derived from COMs tend to lie within the upper range of those obtained from simpler species, as COM transitions are typically optically thinner. Moreover, more robust estimates, relying on optically thin tracers and double-isotopologue techniques, have also been obtained in the literature. In particular, \citet{humire2020} and \citet{yan2023} determined the carbon isotopic ratio using the C$^{34}$S/$^{13}$C$^{34}$S pair toward the +50~km~s$^{-1}$ cloud and along the Sgr~B2 line of sight, respectively, finding values in the range 22--35.

Finally, isotopic fractionation processes in gas-phase chemistry may also contribute to lowering the \cratio\;in molecular species and should therefore be taken into account (e.g., \citealt{colzi2020}). Isotopic fractionation refers to the enrichment or depletion of specific isotopes--such as $^{13}\mathrm{C}$ relative to $^{12}\mathrm{C}$--within molecules due to small differences in reaction rates during isotopic exchange reactions. These effects arise from differences in zero-point energy (ZPE) between isotopologues, leading to temperature-dependent fractionation in which the heavier isotope is preferentially incorporated into the more stable molecular form (e.g., \citealt{caselliceccarelli2012,roueff2015}).

Several recent studies have indicated that the metallicity gradient in the inner Milky Way may flatten as one approaches the Galactic center (e.g., \citealt{ryde2015,kovtyukh2019,feldmeier-krause2020,chen2023,matsunaga2023}). 
In view of this, investigations of isotopic ratios, such as $^{12}$C/$^{13}$C, in molecular gas within the CMZ become especially timely, since isotopic signatures offer a complementary probe of the Galactic chemical evolution under different metallicity conditions and different timescales (e.g., \citealt{grieco2015,romano2017}).
In our work, we aim to determine the initial \cratio\;ratio of the material from which G+0.693 formed, which reflects the stellar nucleosynthetic enrichment history of the CMZ. 
To this end, we use observations of single and double isotopologues of HC$_3$N and HC$_5$N toward the G+0.693$-$0.027 molecular cloud (hereafter G+0.693), identifying those species with negligible line opacity, and combine the derived isotopic ratios with astrochemical models to determine the degree of isotopic fractionation.

This work represents the first paper in a series aimed at building a comprehensive inventory of isotopic ratios in the CMZ source G+0.693. These studies will provide a direct link with isotopic ratio measurements in extragalactic sources, which are now becoming accessible, e.g., through the recent ALMA large observational programsALCHEMI (The ALMA Comprehensive High-resolution Extragalactic Molecular Inventory) toward the nearby starburst galaxy NGC 253 \citep[e.g.][]{martin2021,butterworth2024}.
The paper is organized as follows. In Sect.~\ref{source-observations}, we describe the target source and the observational data used in this study. The analysis procedures and the main observational results are presented in Sect.~\ref{sec:analysis-results}. The fractionation degree is then investigated using astrochemical models in Sect.~\ref{sec:astrochemistry}. The implications of the main results are discussed in Sect.~\ref{discussion}, and our conclusions are summarized in Sect.~\ref{conclusions}.

\section{Source and observations}
\label{source-observations}

In this Section, we describe the main physical and astrochemical properties of the G+0.693 molecular cloud and summarize the observational datasets employed in this study.

\subsection{The G+0.693–0.027 CMZ molecular cloud}
\label{sec-source}

\begin{table}
\centering
\caption{\label{table-physprop} Properties of the two velocity components toward G+0.693 studied in this work, as found by \citet{colzi2024}.}
\begin{tabular}{lcccc}
\hline\hline
Component & RA(ICRS) &  Dec(ICRS) &$n_{\rm H_{2}}$	& $T_{\rm kin}$ \\
& (h m s) & ($\degree\;\prime\;\prime\prime$)& (cm$^{-3}$) & (K) \\
\hline
C1\tablefootmark{a} & 17:47:22.00 & -28:21:27.00 & 2$\times$10$^{4}$ & 140  \\
C2 & 17:47:21.83   &    -28:21:20.57& 5$\times$10$^{4}$ & 30 \\
\hline
\normalsize
\end{tabular}
\vspace{-0.6cm} 
\tablefoot{\tablefoottext{a}{The observations described in this paper are centered on these coordinates.}
}
 \end{table}

G+0.693 is located in the Sgr~B2 region. This cloud is found to be extremely prolific from an astrochemical point of view, 
as many molecules--including numerous COMs--have been detected, some of them for the first time 
(\citealt{zeng2018,rivilla2019,jimenez-serra2020,rivilla2021a,rivilla2023,rodriguez-almeida2021a,jimenez-serra2022,sanz-novo2023,zeng2023,sanandres2024,sanz-novo2025,araki2026}). 

G+0.693 lies approximately $55''$ northeast of Sgr~B2(N), a region well known for its intense ongoing star formation (e.g., \citealt{schmiedeke2016,ginsburg2018}). In contrast, G+0.693 shows no signposts of current high-mass star formation activity, such as UC~HII regions, H$_{2}$O or Class~II methanol masers, or strong dust continuum sources 
(e.g., \citealt{ginsburg2018,zeng2020}). 
\citet{zeng2020} proposed that G+0.693 is currently affected by a cloud--cloud collision, producing the shocks responsible for its rich chemistry (e.g., \citealt{requena-torres2006,martin2008,armijos-abendano2020}). Recent observations toward G+0.693 reveal a broad line component associated with an extended, warm, and diffuse molecular gas structure (C1), as well as two denser and colder condensations (C2 and C3) embedded within it. These condensations appear to be in a post-shock evolutionary stage and are characterized by enhanced deuteration (high D/H ratios) \citep{colzi2022a,colzi2024}. These results suggest that G+0.693 may represent the prestellar precursor of a future massive star-forming cluster within the Sgr~B2 complex.
Altogether, these results highlight G+0.693 as a unique laboratory for investigating the interplay between shocks, potential sites of future star formation, and molecular complexity in the CMZ.

In this work, we focus on the study of carbon isotopic ratios for the C1 and C2 line components analised by \citet{colzi2024}. The coordinates and physical properties for these two gas components are listed in Table~\ref{table-physprop}.

\subsection{Observations}

The work presented here makes use of an ultra-high-sensitivity spectral survey toward the G+0.693 molecular cloud, carried out with the IRAM\footnote{Institut de Radioastronomie Millimétrique.} 30m (Granada, Spain) and Yebes 40m (Guadalajara, Spain) radio telescopes. These datasets have been previously used in several studies (e.g., \citealt{rivilla2021a,colzi2022a,jimenez-serra2022,rivilla2022b,sanz-novo2023,colzi2024}). The observations were performed in position-switching mode, centered at $\alpha_{\rm ICRS}$ = 17$^{\rm h}$47$^{\rm m}$22$^{\rm s}$ and $\delta_{\rm ICRS}$ = $-$28$^\circ$21$^{\prime}$27\asec, with an off position offset by ($-$885\asec, +290\asec). The spectral line intensities are expressed in terms of the antenna temperature, $T_{\rm A}^{*}$, corrected for atmospheric attenuation, radiative losses, and rearward scattering and spillover. This approximation is appropriate given the extended molecular emission observed toward G+0.693, which extends well beyond the primary beams of the telescopes (e.g., \citealt{brunken2010,jones2012,li2020,zheng2024}).

The IRAM 30m and Yebes 40m observations used in this work were obtained from ultra-deep spectral surveys reaching sub-mK noise levels. The IRAM 30m data combine observations from projects 172--18 (PI: Martín-Pintado; April 2019), 018--19 and 133--19 (PI: Rivilla; August and December 2019), and 123--22 (PI: Jiménez-Serra; February 2023). These cover the frequency ranges 71.76--116.72~GHz, 124.77--175.5~GHz, 199.8--238.29~GHz, 252.52--260.30~GHz, and 268.2--275.98~GHz. The spectra were smoothed to 615~kHz, corresponding to a velocity resolution of 1.0--2.2~km~s$^{-1}$, adequate to resolve the typical line widths of 15--20~km~s$^{-1}$ observed in the region (e.g., \citealt{rivilla2020,rivilla2021a,colzi2022a,colzi2024}). The achieved root mean square (rms) noise levels are 1.7--2.8~mK (71--90~GHz), 1.5--9.8~mK (90--115~GHz), 3.1--6.8~mK (124--175~GHz), 4.7--9.7~mK (200--238~GHz), and 10.8--18~mK (250--275~GHz). The beam varies from 10\asec\;up to 32\asec\;within the frequencies covered.

The Yebes 40m observations were obtained between March 2021 and March 2022 (project 21A014; PI: Rivilla), covering the 31.07--50.42~GHz range, with a total on-source time of 110~hr\footnote{Individual observing sessions were performed between 2021 March and 2022 March; see \citet{rivilla2023} and \citet{sanz-novo2023} for details.}. The final spectra were smoothed to 256~kHz (velocity resolution of 1.5--2.5~km~s$^{-1}$), yielding rms noise levels of 0.25--0.9~mK across the band. The beam varies from 36.4\asec\;up to 54.4\asec\;within the frequencies covered.

\section{Analysis and results}

\subsection{Observational results}
\label{sec:analysis-results}

\begin{figure*}[h]
\centering
\includegraphics[width=\textwidth]{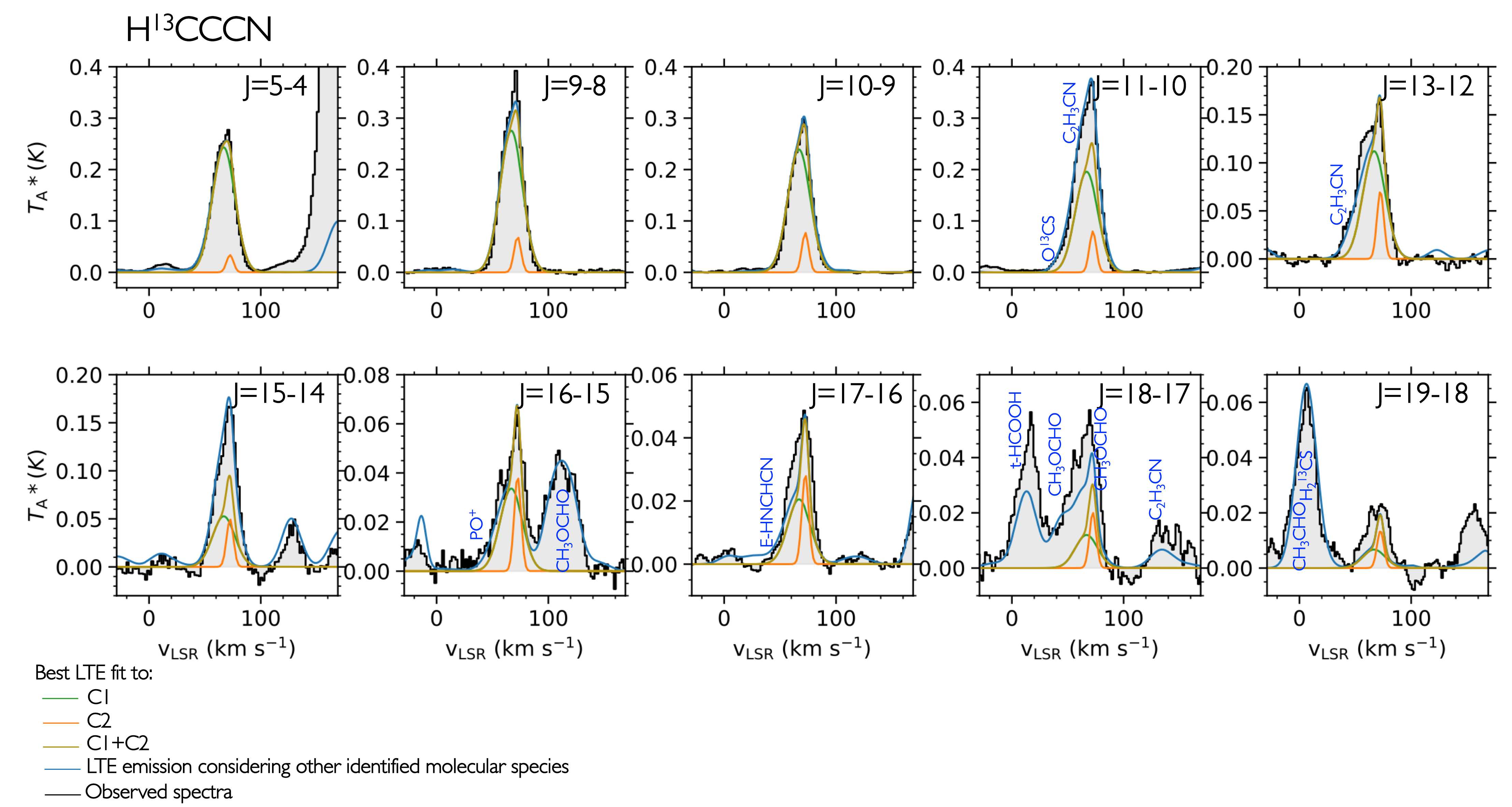}
\caption{H$^{13}$CCCN observed transitions used for the LTE analysis (black histrogram). For each panel, the corresponding transition is indicated in the upper right corner. The green and orange solid lines are the best LTE fit to the C1, and C2 components, respectively. The dark gold line is the sum of the two components. The blue line indicates the total modeled line emission, including also the contribution of all molecular species previously identified in the survey.} 
\label{fig-h13cccn-fit}
\end{figure*}

\begin{figure}[h]
\centering
\includegraphics[width=\columnwidth]{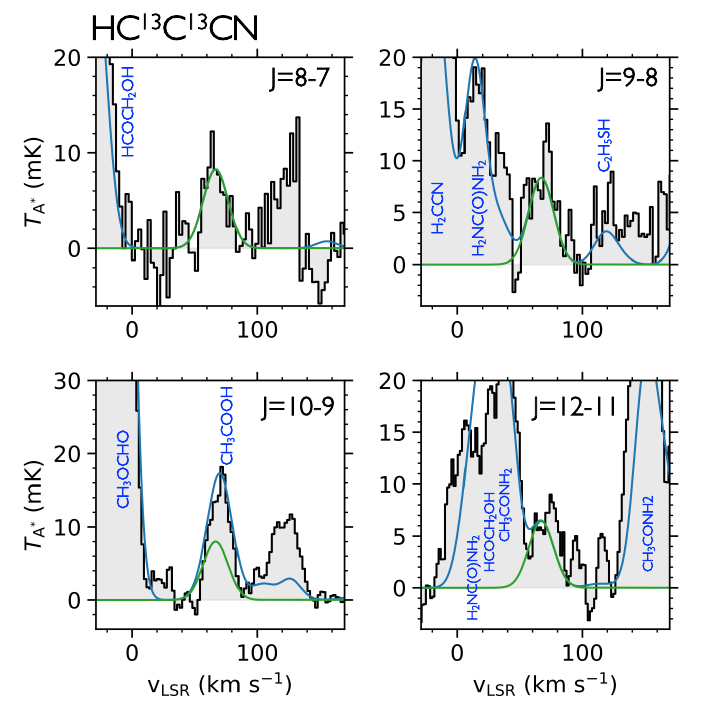}
\caption{Same as Fig.~\ref{fig-h13cccn-fit} but for HC$^{13}$C$^{13}$CN. In this case only the C1 component modeled in green is detected, as explained in Sect.~\ref{sec:analysis-results}.}
\label{fig-hc13c13cn-fit}
\end{figure}

\begin{figure*}[h]
\centering
\includegraphics[width=\textwidth]{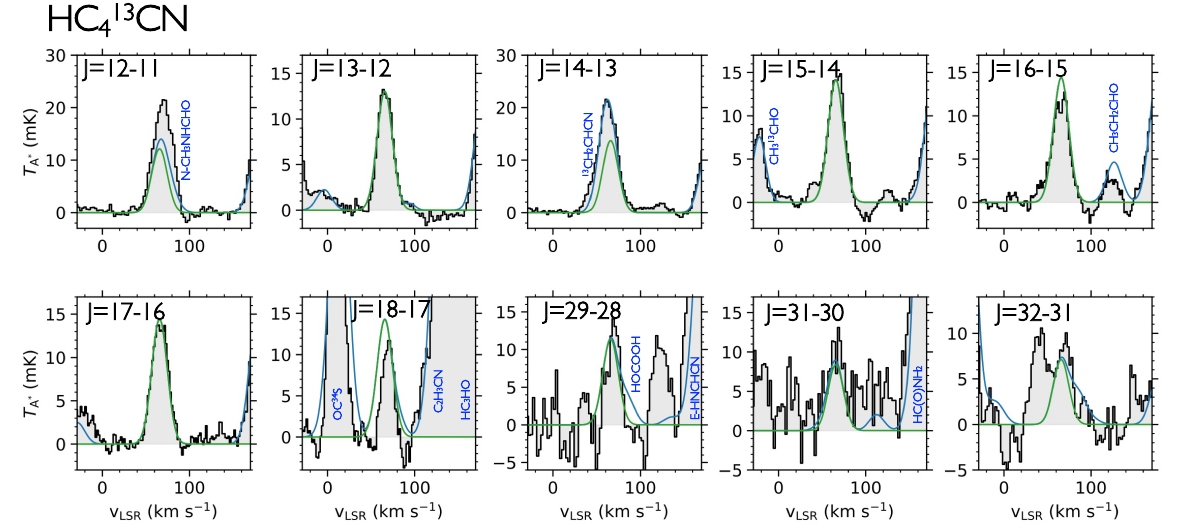}
\caption{Same as Fig.~\ref{fig-hc13c13cn-fit} but for HC$_{4}$$^{13}$CN.}
\label{fig-hc413cn-fit}
\end{figure*}

\begin{figure*}[h]
\centering
\includegraphics[width=\textwidth]{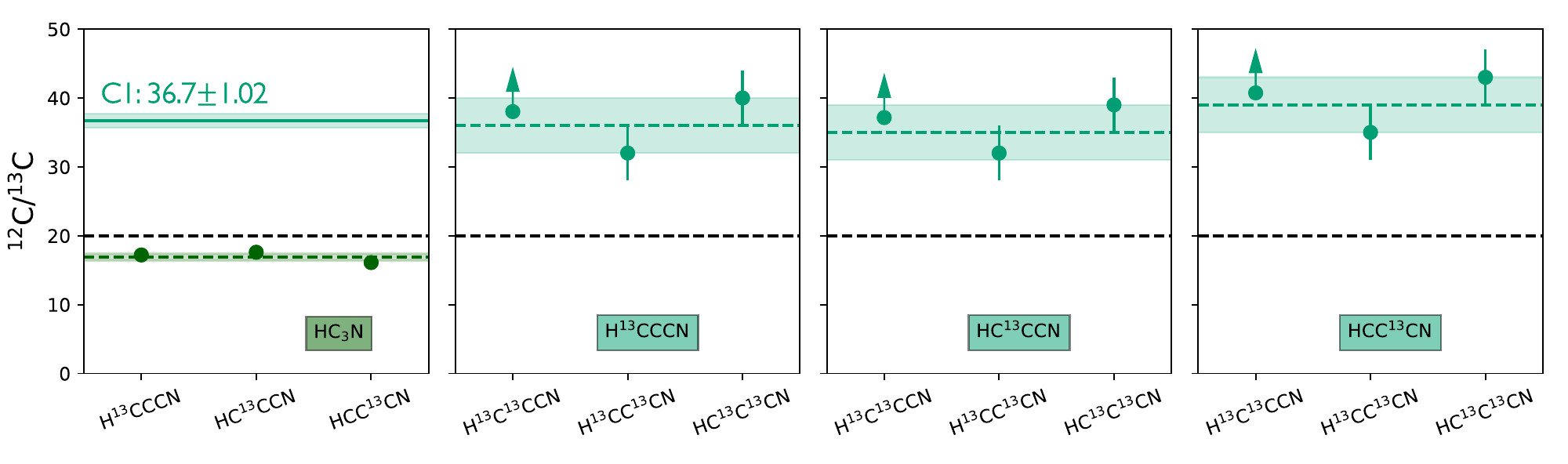}
\caption{Carbon isotopic ratios ($^{12}$C/$^{13}$C) measured for different isotopologues of HC$_3$N toward G+0.693 (C1 component) in the Galactic Center. From left to right, the panels show the ratios obtained with HC$_3$N, H$^{13}$CCCN, HC$^{13}$CCN, and HCC$^{13}$CN in the numerator, respectively. Black dashed horizontal lines indicate the canonical Galactic Center value of $\sim$20. Points with arrows denote lower limits, and scatter points with error bars represent individual measurements. Light green corresponds to ratios derived from double-isotopologues, while olive green corresponds to ratios from the main isotopologue. In each panel, the dashed green line and shaded area indicate the average value and uncertainty for that molecule. Moreover, in the first panel, the solid light green line and shaded area shows the average ratio derived from the double isotopologues.}
\label{fig-cratiohc3n-c1}
\end{figure*}

\begin{figure*}[h!]
\centering
\includegraphics[width=\textwidth]{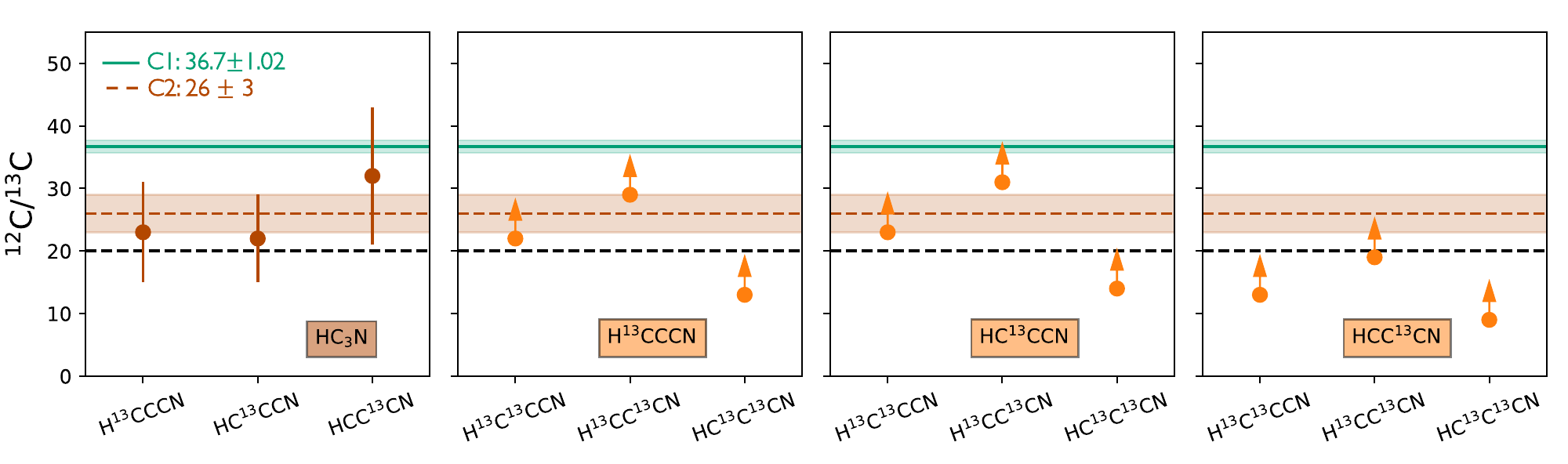}
\caption{Same as Fig.~\ref{fig-cratiohc3n-c1}, but for C2. Black dashed horizontal lines indicate the canonical Galactic Center value of $\sim$20. Light orange arrows indicate the lower limits on the ratios derived from the double isotopologues, while dark orange corresponds to those derived from the main isotopologue. The dashed dark orange line and shaded region in all panels represent the average value and its uncertainty derived from the measurements shown in the first panel. The average value obtained for C1 is also shown as a solid light green line with its corresponding shaded region for the error (as in the first panel of Fig.~\ref{fig-cratiohc3n-c1}).}
\label{fig-cratiohc3n-c2}
\end{figure*}

\begin{figure}[h!]
\centering
\includegraphics[width=\columnwidth]{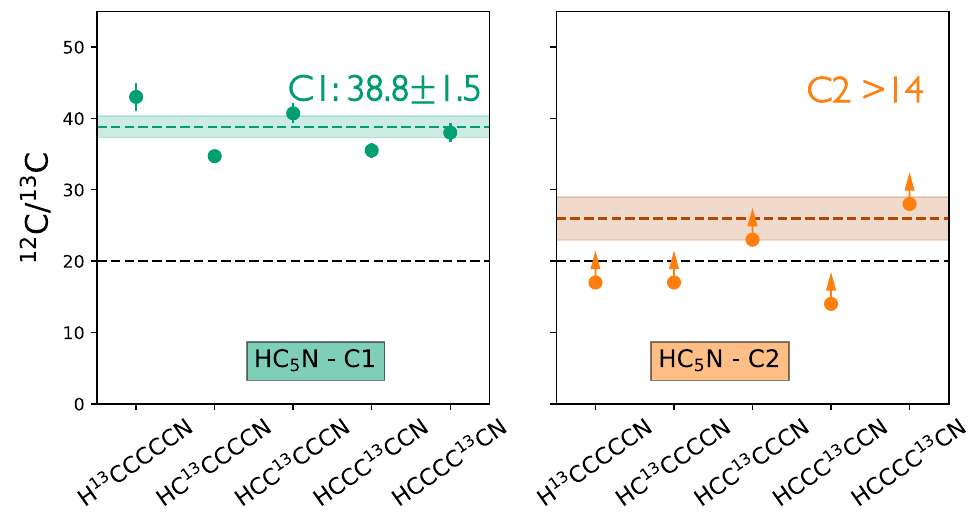}
\caption{$^{12}$C/$^{13}$C ratios derived from HC$_5$N for the C1 component (left panel) and the C2 component (right panel). In both panels the black dashed horizontal lines indicate the canonical Galactic Center value of $\sim$20. In the left panel, the light green dashed line and shaded region indicate the average value and its uncertainty derived from HC$_5$N. In the right panel, the dark orange line and shaded region show the average value obtained for C2 from HC$_3$N, illustrating that the lower limits derived from HC$_5$N are consistent with this value.}
\label{fig-cratiohc5n}
\end{figure}

\begin{table}
\centering
\caption{Column densities and isotopic ratios for \3n.}
\label{table-ratios-hc3n}
\setlength{\tabcolsep}{2pt}
\begin{tabularx}{0.94\columnwidth}{l c r r}
\hline\hline
Molecule & Comp. & $N$ & \cratio \\
 & & ($\times10^{14}$ cm$^{-2}$) & \\
\hline
HC$_3$N\tablefootmark{a} & C1 & 6.54$\pm$0.07\tablefootmark{g} & \\
                & C2 & 9$\pm$3 & \\
\hline
H$^{13}$CCCN & C1 & 0.380$\pm$0.004 & 17.2$\pm$0.3 \\
             & C2 & 0.39$\pm$0.04 & 23$\pm$8 \\
HC$^{13}$CCN & C1 & 0.372$\pm$0.007 & 17.6$\pm$0.4 \\
             & C2 & 0.41$\pm$0.05 & 22$\pm$7\\
HCC$^{13}$CN & C1 & 0.407$\pm$0.006 & 16.1$\pm$0.3 \\
             & C2 & 0.27$\pm$0.04 & 32$\pm$11 \\
\hline
H$^{13}$C$^{13}$CCN & C1 & $<$0.01 & $>$38 / $>$37 / $>$40\tablefootmark{b}\\
                    & C2 & $<$0.018\tablefootmark{c} & $>$22 / $>$23\tablefootmark{d} \\
                    & C2 & $<$0.022\tablefootmark{e} & $>$13\tablefootmark{f} \\
H$^{13}$CC$^{13}$CN & C1 & 0.0117$\pm$0.0014 & 32$\pm$4 / 32$\pm$4 / 35$\pm$4\tablefootmark{b} \\
                    & C2 & $<$0.013\tablefootmark{c} & $>$28 / $>$30\tablefootmark{d} \\
                    & C2 & $<$0.015\tablefootmark{e} & $>$19\tablefootmark{f} \\
HC$^{13}$C$^{13}$CN & C1 & 0.0095$\pm$0.0010 & 40$\pm$4 / 39$\pm$4 / 43$\pm$4\tablefootmark{b} \\
                    & C2 & $<$0.029\tablefootmark{c} & $>$13 / $>$14\tablefootmark{d} \\
                    & C2 & $<$0.032\tablefootmark{e} & $>$8\tablefootmark{f} \\
\hline
\multicolumn{4}{c}{Average values C1\tablefootmark{h}}\\
\hline
\multicolumn{3}{l}{HC$_3$N/single $^{13}$C} &16.9$\pm$0.5 \\
\multicolumn{3}{l}{H$^{13}$CCCN/double $^{13}$C} &36$\pm$4 \\
\multicolumn{3}{l}{HC$^{13}$CCN/double $^{13}$C} &35$\pm$4 \\
\multicolumn{3}{l}{HCC$^{13}$CN/double $^{13}$C} &39$\pm$4 \\
\multicolumn{3}{l}{Mean of double isotopologues} &36.7$\pm$1.02 \\
\hline
\multicolumn{4}{c}{Average value C2}\\
\hline
\multicolumn{3}{l}{HC$_3$N/single $^{13}$C} &26$\pm$3 \\
\hline
\end{tabularx}
\tablefoot{
\tablefoottext{a}{Taken from \citet{colzi2024}. Only the low-$J$ transitions are included in the results of this paper.}
\tablefoottext{b}{Relative to H$^{13}$CCCN / HC$^{13}$CCN / HCC$^{13}$CN.}
\tablefoottext{c}{Calculated using the same $T_{\rm ex}$ of the single $^{13}$C isotopologues H$^{13}$CCCN and HC$^{13}$CCN (see Table \ref{table-fit}).}
\tablefoottext{d}{Relative to H$^{13}$CCCN / HC$^{13}$CCN.}
\tablefoottext{c}{Calculated using the same $T_{\rm ex}$ of the single $^{13}$C isotopologue HCC$^{13}$CN (see Table \ref{table-fit}).}
\tablefoottext{f}{Relative to HCC$^{13}$CN.}
\tablefoottext{g}{The fractional abundance of HC$_3$N relative to H$_2$ is $(4.8\pm0.7)\times10^{-9}$, adopting an H$_2$ column density of $N(\mathrm{H}_2)=1.35\times10^{23}$ cm$^{-2}$ from \citet{martin2008}, assuming an uncertainty of 15\% of its value.}
\tablefoottext{h}{Average values are calculated as the arithmetic mean of the measurements \(\bar{x}=(1/N)\sum_{i=1}^{N}x_i\), where \(x_i\) is the \(i\)-th measurement and \(N\) is the number of measurements, while their uncertainties $\sigma_{\bar{x}}$ are estimated from the data standard deviation  \( s = \sqrt{\frac{1}{N-1}\sum_{i=1}^{N}(x_i-\bar{x})^2} \rightarrow  
 \sigma_{\bar{x}} = s/\sqrt{N}. \)}
}
\end{table}

\begin{table}
\centering
\caption{Column densities and isotopic ratios for \5n.}
\label{table-ratios-hc5n}
\begin{tabularx}{0.87\columnwidth}{l c r r}
\hline\hline
Molecule & Comp. & $N$ & \cratio \\
 & & ($\times10^{14}$ cm$^{-2}$) & \\
 \hline
HC$_5$N\tablefootmark{b}& C1 & 1.247$\pm$0.018\tablefootmark{a} & \\
                & C2 & 1.7$\pm$0.3 & \\
\hline
H$^{13}$CC$_4$N & C1 & 0.0288$\pm$0.0013 & 43$\pm$2 \\
                & C2 & $<$0.098 & $>$17 \\
HC$^{13}$CC$_3$N & C1 & 0.0355$\pm$0.0005 & 35.1$\pm$0.7 \\
                 & C2 & $<$0.098 & $>$17 \\
HC$_2^{13}$CC$_2$N & C1 & 0.0302$\pm$0.0010 & 41.2$\pm$1.4 \\
                   & C2 & $<$0.075 & $>$23 \\
HC$_3^{13}$CCN & C1 & 0.0347$\pm$0.0010 & 35.9$\pm$1.1 \\
                & C2 & $<$0.124 & $>$14 \\
HC$_4^{13}$CN & C1 & 0.0324$\pm$0.0010 & 38.5$\pm$1.3 \\
               & C2 & $<$0.060 & $>$28 \\
\hline
\multicolumn{4}{c}{Average value C1}\\
\hline
\multicolumn{3}{l}{HC$_5$N/single $^{13}$C} &38.8$\pm$1.5 \\
\hline
\end{tabularx}
\tablefoot{
\tablefoottext{a}{The fractional abundance of HC$_5$N relative to H$_2$ is $(9.2\pm1.4)\times10^{-10}$, adopting an H$_2$ column density of $N(\mathrm{H}_2)=1.35\times10^{23}$ cm$^{-2}$ from \citet{martin2008}, assuming an uncertainty of 15\% of its value.}
\tablefoottext{b}{Only the low-$J$ transitions are included in the results of this paper.}
}
\end{table}

To reliably derive the \cratio\;isotopic ratio of HC$_3$N and HC$_5$N, we searched for all possible single- and double-$^{13}$C isotopologues of these species. 
Figures~\ref{fig-h13cccn-fit}--\ref{fig-hc413cn-fit} and \ref{fig-hc13ccn-fit}--\ref{fig-hc213cc2n-fit} show the rotational transitions of H$^{13}$CCCN, HC$^{13}$C$^{13}$CN,  HC$_4^{13}$CN, HC$^{13}$CCN, HCC$^{13}$CN, H$^{13}$C$^{13}$CCN, H$^{13}$CC$^{13}$CN, H$^{13}$CC$_4$N, HC$^{13}$CC$_3$N, HC$_3^{13}$CCN, and HC$_2^{13}$CC$_2$N detected toward G+0.693. In addition, Appendix~\ref{app-hc5n} presents the HC$_5$N transitions covered in our data. Their spectroscopic information is taken from the Cologne Database for Molecular
Spectroscopy\footnote{\url{http://cdms.astro.uni-koeln.de/classic/}} (CDMS; \citealt{muller2001,muller2005}; \citealt{endres2016}). The entries of the catalogue for these species are based on the laboratory works of \citet{dezafra1971,alexander1976,creswell1977,mallinson1978,winnewisser1982,chen1991,yamada1995,thorwirth2000,thorwirth2001,bizzocchi2004,giesen2020b,sanz2005}. The dipole moments were determined by \citet{alexander1976,deleon1985}. 
Note that we do not display and use in the analysis the rotational transitions that are heavily blended with other molecular species or whose emission lies below the noise level. Only in a few cases--such as H$^{13}$C$^{13}$CCN, discussed below--these transitions are shown since they provide further support for the derived model fits. In fact, some transitions are blended with lines from other species, but their predicted intensities remain useful for reproducing the observed spectra once the contributions of all species previously identified toward G+0.693 are included. Furthermore, the hyperfine structure (HFS) of the molecular transitions is neglected in our line profile modeling. In our observations, the maximum frequency separation due to HFS occurs in the $\text{HC}_3\text{N} (J = 5-4)$ transition, where the $F = 4 \to 3$ component is offset from the $F = 6 \to 5$ component by $0.2494$ MHz. This corresponds to a velocity separation of $\approx 1.6 \text{ km s}^{-1}$, which is negligible given the broad observed line widths ($\text{FWHM} \approx 20 \text{ km s}^{-1}$). Since HFS splitting scales inversely with the rotational quantum number $J$, these components remain unresolved and do not significantly contribute to the observed line profiles.

To derive the total column densities of all molecules, we performed a local thermodynamic equilibrium (LTE) analysis. 
For HC$_5$N, we carried out a dedicated analysis following the same methodology applied to HC$_3$N in \citet{colzi2024}. 
The full set of observed transitions cannot be simultaneously reproduced under LTE conditions. In fact, possible non-LTE effects introduce a curvature in the rotational diagram \citep{goldsmith1999}, as illustrated in Fig.~2 of \citet{colzi2024}.
Thus, as in the previous work, we first excluded all transitions with upper-level energies $E_{\rm up}<5$\,K ($J=5$--4 and 7--6), whose observed intensities exceed the predictions of LTE models. 
These two low-energy transitions of \5n\;are therefore not considered further in the analysis. 
The remaining transitions were separated into two groups:  
(i) 5\,$<E_{\rm up}\leq$\,70\,K (“low-$J$”, from $J=12$--11 to 32--31; Fig.~\ref{fig-hc5n-low}), and  
(ii) $E_{\rm up}>$\,70\,K (“high-$J$”, from $J=33$--32 to 43--42; Fig.~\ref{fig-hc5n-high}). Transitions at even higher $J$ are too faint to be detected in our data.
Note that for HC$_{3}$N and its isotopologues the low-$J$ regime is defined for 5\,$<E_{\rm up}\leq$\,100\,K (see also \citealt{colzi2024}). For HC$_{5}$N we have redefined this in order to cover the same energy range as its $^{13}$C-isotopologues, for which only transitions with $E_{\rm up}<$\,70\,K have been detected. We verified that extending the HC$_{5}$N analysis up to $E_{\rm up}$=100 K (i.e. including transitions up to $J$=39--38) does not significantly affect the derived total column density, yielding consistent values within the uncertainties ((1.248$\pm$0.011)$\times$10$^{14}$ cm$^{-2}$ and (1.247$\pm$0.018)$\times$10$^{14}$ cm$^{-2}$).

For the $^{13}$C isotopologues of HC$_{3}$N and HC$_{5}$N studied here, only transitions corresponding to the"low-$J$" regime of the main isotopologues are detected, as the "high-$J$" lines remain below the noise level.  
Thus, all isotopic ratios discussed in the following are derived exclusively from these "low-$J$" transitions. Finally, we note that the C1 velocity component is detected in all species analyzed, whereas the C2 component is detected only for the main isotopologues HC$_{3}$N and HC$_{5}$N, and for the single $^{13}$C-isotopologues of HC$_{3}$N.

To fit the groups of transitions under LTE conditions, we used the Spectral Line Identification and Modeling (\texttt{SLIM}) tool (version from 2024, June 15) within the \texttt{MADCUBA} software package, which generates synthetic spectra assuming LTE and taking line opacity into account. 
For each transition, \texttt{SLIM} computes the convolution of a Gaussian source distribution with the corresponding Gaussian telescope beam; further details are provided in Sect.~3.2.2 of \citet{martin2019}. 
The physical parameters describing the molecular emission were obtained using the automatic fitting routine \texttt{AUTOFIT}, which performs a nonlinear least-squares LTE optimization based on the Levenberg–Marquardt algorithm \citep{martin2019}.

The free parameters considered in the \texttt{MADCUBA} fit are the column density of the molecule ($N$), the excitation temperature ($T_{\rm ex}$), the velocity ($\varv_{\rm LSR}$), the full width at half maximum (FWHM) and the source size ($\theta_{\rm source}$). 
As determined in \citet{colzi2024}, we considered component C1 to be extended for the line fitting, meaning that no beam dilution was applied, while for C2, we used a source size\footnote{This explains the diverging evolution of the modeled line intensities for the C1 and C2 components with increasing $J$ observed in Figs.~\ref{fig-h13cccn-fit} and \ref{fig-hc13ccn-fit}. Specifically, the intensities of higher $J$ transitions become more prominent as the coupling between the source size and the telescope beam improves at higher frequencies.} of 9\asec. The FWHM of the C1 component of \5n\;has been left as a free parameter (as for \3n), while for C2 we assumed the same value as \3n\;to ensure convergence of the fit. For the $^{13}$C-isotopologues, the FWHM has been fixed to that of the same component of the main isotopologue. $T_{\rm ex}$ and $\varv_{\rm LSR}$ were left free in the fitting procedure unless their relative errors were higher than 15\%. In the latter case, these parameters were also fixed to the values of the other isotopologues used to derive the \cratio\;ratio or, when possible, to the average value of similar species.
The final column densities were derived using the antenna temperature ($T_{\rm A}^{*}$) scale for the C1 component and the main beam temperature ($T_{\rm MB}$) scale for the C2 component, which has been found to be a compact condensation (\citealt{colzi2024}). $T_{\rm MB}$ was obtained from $T_{\rm A}^{*}$ using the formula $T_{\rm A}^{*} = T_{\rm MB}\,\eta_{\rm MB}$, where $\eta_{\rm MB}$ is the ratio of the main beam efficiency, $B_{\rm eff}$, and the forward efficiency of the telescope\footnote{The beam efficiencies of the telescopes are available online at 
\url{https://publicwiki.iram.es/Iram30mEfficiencies} and 
\url{https://rt40m.oan.es/rt40m_en.php} for the IRAM 30m and Yebes 40m telescopes, respectively.}, $F_{\rm eff}$.

When one of the isotopologues was not detected, we derived an upper limit to its column density. These limits were estimated using $3\times{\rm rms}$ at the corresponding rest frequency, together with the assumed excitation temperature ($T_{\rm ex}$) and linewidth (FWHM) (see Table \ref{table-fit}). For the C2 component of the double $^{13}$C isotopologues we computed several upper limits because the adopted $T_{\rm ex}$ depends on which single $^{13}$C isotopologue is taken as reference. This procedure ensures that the resulting lower limits on the \cratio\;isotopic ratios are derived consistently and accurately. The only exception is the column density derived for H$^{13}$C$^{13}$CCN (Fig.~\ref{fig-h13c13ccn-fit}). Two of its transitions are slightly blended with emission from other species, preventing a reliable fit. For this reason, we provide an upper limit obtained by manually increasing $N$ until the synthetic spectrum reproduced the full observed line profile.
The best LTE fits are shown superimposed on the observed spectra in Figs.~\ref{fig-h13cccn-fit}--\ref{fig-hc413cn-fit}, \ref{fig-hc13ccn-fit}--\ref{fig-hc213cc2n-fit}, \ref{fig-hc5n-low}, and \ref{fig-hc5n-high}. The corresponding fitting results are summarized in Table~\ref{table-fit}.

From the derived total column densities, we calculated the \cratio\ ratios in two ways: (i) by dividing the column density of the main isotopologue by those of the single $^{13}$C isotopologues (for both \3n\ and \5n), and (ii) for HC$_3$N only, by dividing the single $^{13}$C isotopologues by the double $^{13}$C isotopologues, considering all possible combinations. 
The resulting isotopic ratios are reported in Tables~\ref{table-ratios-hc3n} and \ref{table-ratios-hc5n}, for HC$_3$N and HC$_5$N respectively, and are graphically presented in Figs.~\ref{fig-cratiohc3n-c1}, \ref{fig-cratiohc3n-c2}, and \ref{fig-cratiohc5n}.

For the C1 component of \3n, the \cratio\;ratios derived from the main isotopologue are approximately factor of two lower than those obtained from double–$^{13}$C species. This discrepancy likely stems from moderate optical depth in the \3n\;lines ($\tau \sim 0.1$--$0.5$) and a lack of sufficiently optically thin transitions to constrain the opacity, which is a lower limit. Consequently, also the column densities derived from the main isotopologue should be treated as lower limits. In fact, the optically thinnest lines are all in the high-$J$ group, where $\tau$ ranges only from 0.001 to 0.04. Furthermore, we have assumed that the emitting gas is more extended than the telescope beam; if this is not the case, the derived opacities would represent lower limits. It should be noted that, if the HC$_{3}$N column density of the C1 component is increased to reproduce the \cratio\;ratio derived from the double isotopologues, a source size of $\gtrsim$25--30\asec\;would be required to match the observed spectra. In this case $\tau$ would be better constrained, reaching values up to 2--3. In contrast, the source size is constrained for the C2 component, and even though some transitions are optically thick, we were able to apply reliable opacity corrections to the final column densities (with $\tau$ between 0.1 and 2.5). An additional indication of optical–depth effects in C1 is that the $T_{\rm ex}$ derived for the main isotopologue is lower--and inconsistent with--the excitation temperature of the optically thin $^{13}$C isotopologues, whereas in C2 the $T_{\rm ex}$ values agree within the uncertainties.
Based on these considerations, we conclude that the most reliable isotopic ratio for the C1 component of \3n\;is that derived from the double $^{13}$C isotopologues, which yields an average value of 36.7$\pm$1.02. For the C2 component, the ratio derived from the main isotopologue is considered robust, giving a value of 26$\pm$3. 

For \5n, for which only the C1 component is detected in the $^{13}$C isotopologues, the main isotopologue is already optically thin ($\tau \sim 0.01$--$0.03$). Consequently, the \cratio\;ratios derived from the main species are considered reliable. We also note that, in this case, the excitation temperatures obtained for the main and the $^{13}$C isotopologues are mutually consistent.  
The final isotopic ratio, computed as the average over all possible isotopologue combinations, is 38.8$\pm$1.5.

The average isotopic ratio derived from \3n\;and \5n\;is 37.7$\pm$1.0, which is higher than the values reported from gas–phase molecular emission in the Galactic Center prior to 2020 (grey regions in Figs.~\ref{fig-cratiohc3n-c1}, \ref{fig-cratiohc3n-c2}, and \ref{fig-cratiohc5n}). As we discuss in Sect.~\ref{sec:comparisonothersources}, more recent studies employing methods less affected by the lines opacity of the main isotopologues have obtained ratios consistent with those found here. Altogether, these results indicate that the \cratio\;ratio in the CMZ is likely higher than the commonly adopted value of 20.

\subsection{Astrochemical modeling of isotopologues}
\label{sec:astrochemistry}

\begin{figure*}[h!]
\centering
\includegraphics[width=\textwidth, trim=0cm 0cm 0cm 0cm, clip]{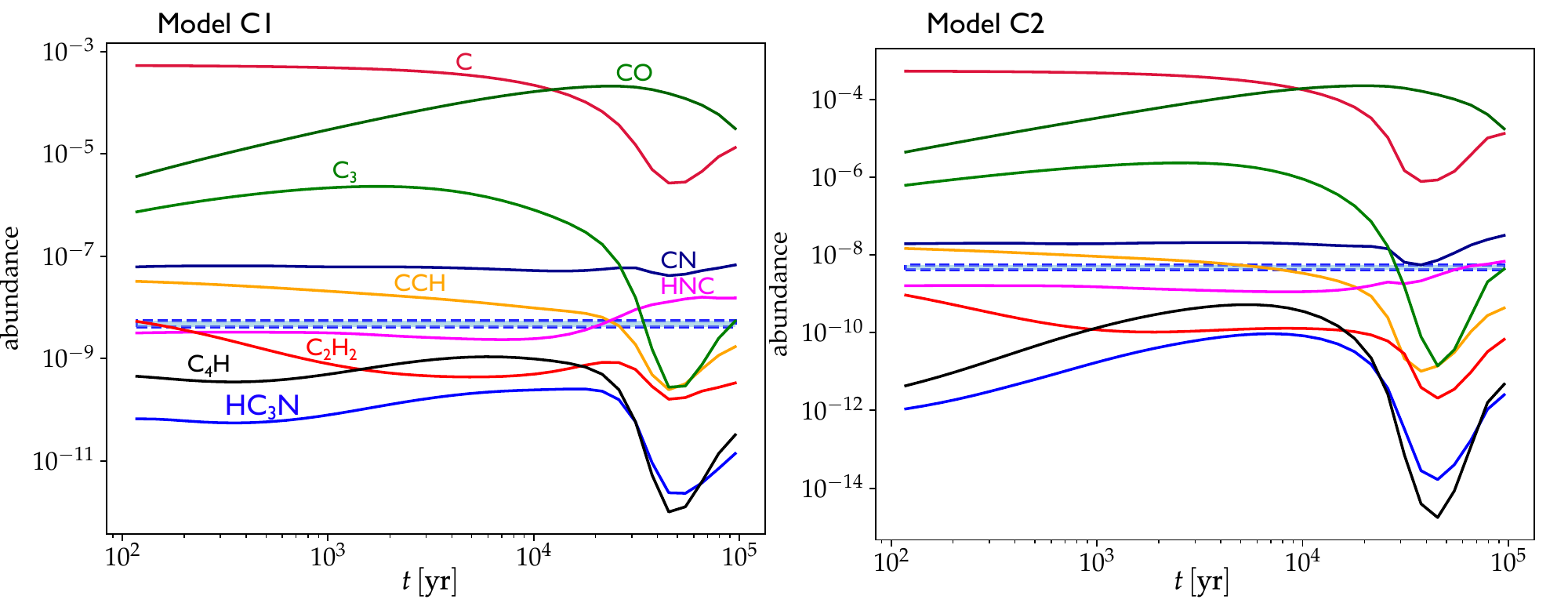}
\caption{Time evolution of the simulated abundances (relative to H$_2$) of HC$_3$N and its main precursor species for the two physical models considered in this work. The left and right panels show the results for the C1 and C2 components, respectively. In both panels the blue horizontal region represents the observed abundance toward the C1 component. The abundances are calculated using the chemical network of \citet{sipila2023} and adopting the physical parameters described in Sect.~\ref{sec:astrochemistry}. The figure highlights that CCH, C, and C$_3$ are among the most abundant carbon-chain precursors at early times, enabling the formation of C$_4$H and subsequently HC$_3$N through the reaction C$_4$H + N. At later times, CN and HNC become increasingly important in the chemistry.}
\label{fig-abundances}
\end{figure*}

\begin{figure*}[h!]
\centering
\includegraphics[width=\textwidth, trim=0cm 0cm 0cm 0cm, clip]{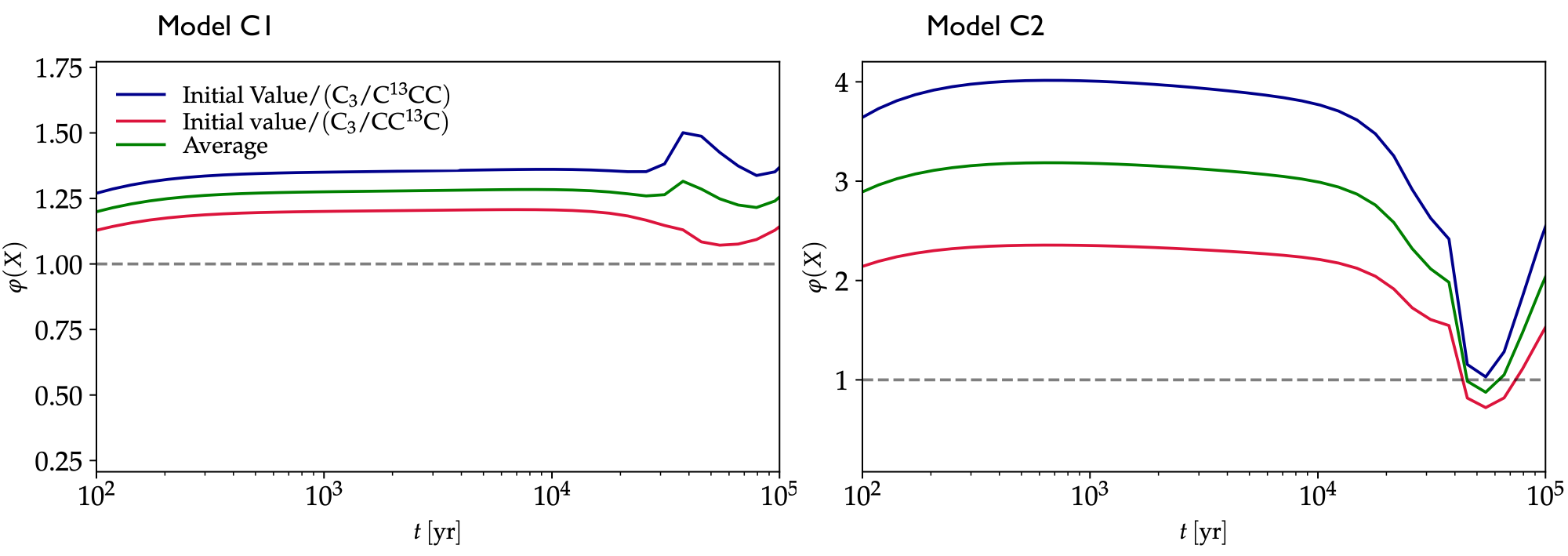}
\caption{Predicted isotopic fractionation degree, $\varphi$(C$_3$), as a function of time for the two models. The quantity $\varphi$(X) is defined as the ratio between the initial elemental $^{12}$C/$^{13}$C ratio and the modeled isotopic ratio for species X. In both panels the black dashed horizontal line represents $\varphi$(X)=1. The left panel shows the results for the model of C1, while the right panel corresponds to the model of C2. The curves represent the fractionation for the different $^{13}$C substitutions in C$_3$, together with their average value. At early times ($\lesssim 3\times10^{4}$ yr), isotopic exchange reactions between $^{13}$C and C$_3$ drive the fractionation degree toward values close to $e^{\Delta E/T}$, leading to moderate fractionation in the C1 model and stronger effects in the lower-temperature C2 model.}
\label{fig-fractionationdegree-C3}
\end{figure*}

\begin{figure*}[h!]
\centering
\includegraphics[width=\textwidth, trim=0cm 0cm 0cm 0cm, clip]{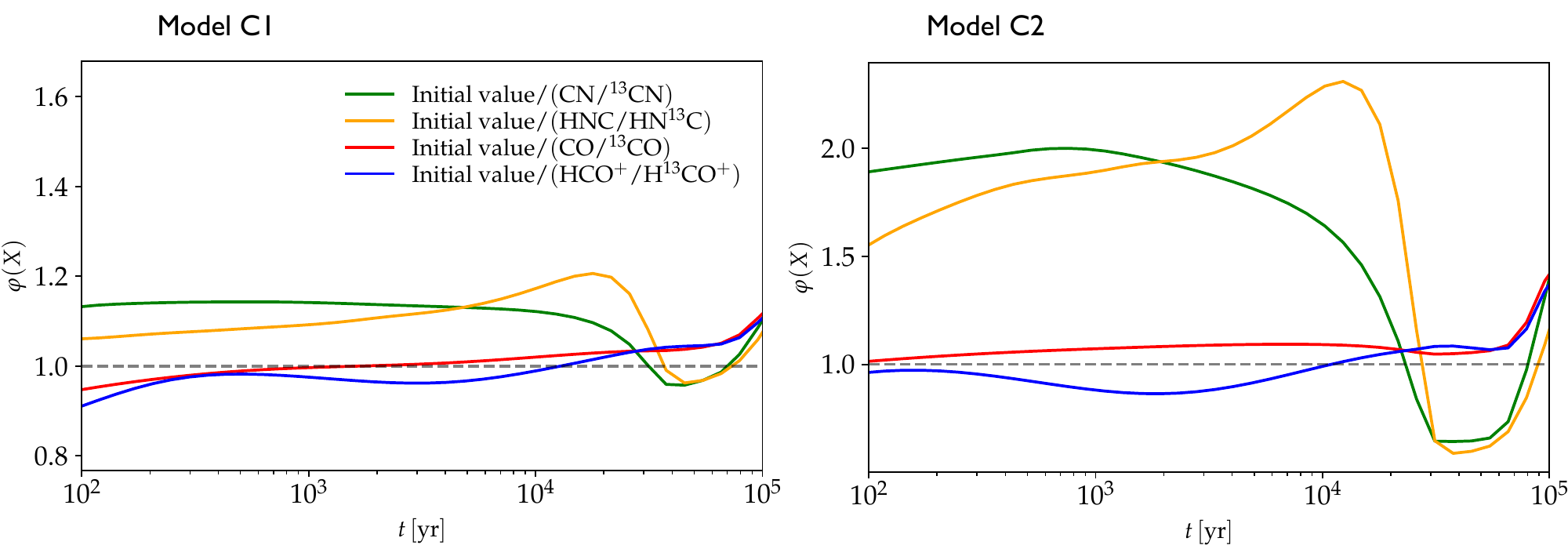}
\caption{Same as Fig.~\ref{fig-fractionationdegree-C3} but for CN, HNC, CO, and HCO$^{+}$ as a function of time for the models of C1 (left) and C2 (right). The fractionation degree decreases below unity at intermediate times, reflecting the role of CO and HCO$^{+}$ as the main sinks of $^{13}$C in the gas phase. At later times, the increasing abundance of He$^{+}$ produced by cosmic rays leads to the reformation of atomic carbon and carbon chains, causing the fractionation degree to rise again toward values close to unity. These two species are therefore expected to be only weakly affected by isotopic fractionation over most of the chemical evolution.}
\label{fig-fractionationdegree-otherspecies}
\end{figure*}

In this section, we investigate the possible effects of isotopic fractionation on the derived \cratio\;ratio. 
We employ the chemical model developed by \citet{sipila2023}, which incorporates an updated chemical network including all combinations of $^{13}$C-, $^{15}$N-, and $^{13}$C$^{15}$N-bearing molecular species, together with the most recent low-temperature isotopic exchange reactions. This chemical network includes $^{13}$C in molecules with up to 3 C atoms. This means that the fractionation of larger precursors of HC$_{3}$N and HC$_{5}$N, like C$_{4}$H, is not treated. Thus, we have first analised the primary chemical pathways responsible for the formation of HC$_{3}$N, and subsequently investigate the \cratio\;ratio of its main precursor species. We have performed two models, for the two gas components C1 and C2, respectively. In particular, we have assumed as gas kinetic temperatures, $T_{\rm kin}$, and H$_{2}$ densities, $n_{\rm H_2}$, those reported in Table \ref{table-physprop}, which remain constant throughout the simulation. We have also assumed a dust temperature, $T_{\rm dust}$, of 20~K, as typically found in the CMZ (e.g., \citealt{rodriguez-fernandez2004,henshaw2023,battersby2025}). Moreover, we have used a cosmic-ray ionization rate of 1.3$\times$10$^{-15}$ s$^{-1}$, as estimated for G+0.693 \citep{rivilla2022b,sanz-novo2024}. Finally, because of the high column densities of gas present in the CMZ, the G+0.693 molecular cloud is highly shielded from external ultraviolet photons, and then we have assumed a visual extinction, $A_{\rm V}$ of 40 mag. The rest of the model parameters are assumed the same as in \citet{sipila2023}. The initial elemental abundances are those used in \citet{jimenez-serra2018}. In the following, we study the chemistry up to 10$^{5}$~yr, as previous works have shown that the molecular abundances observed toward G+0.693 can be reproduced for an evolutionary time between 10$^{4}$ and 10$^{5}$~yr after the shock event (see, e.g., \citealt{requena-torres2006,sanz-novo2024}).

In Fig.~\ref{fig-abundances} we show the simulated abundances of HC$_{3}$N and its main precursors, for the models of C1 and C2, respectively. The behavior in both models is very similar with up to an order of magnitude lower abundances for the low gas-phase temperature model C2. In both models we have found that the main formation pathway to form HC$_{3}$N at almost all time scales, up to $\sim$3$\times$10$^{4}$ yr, is through C$_{4}$H + N (see similar behaviour in abundances in Fig.~\ref{fig-abundances} and scheme in Fig.~\ref{scheme-hc3n}). At 10$^{5}$ yr this reaction is responsible for 43\% of HC$_{3}$N formation, followed by C$_{3}$N$^{-}$ + H (24\%), C$_{2}$H$_{2}$ + CN (10\%), and CCH + HNC (7\%), among other less important reactions.  
In general, C$_4$H is formed in the gas phase starting from the reaction CCH + C $\rightarrow$ C$_3$, followed by subsequent reactions involving C, H, H$_2$, and electrons. CCH, C, and C$_3$ are among the most abundant precursor species up to  $\sim$3$\times$10$^{4}$ yr, whereas at later times HNC and CN become increasingly important. Overall, C$_3$ represents the final key precursor in the chemical formation pathway of HC$_3$N and can therefore be used to investigate the degree of its isotopic fractionation. It should also be noted that at these later timescales the main sink of atomic carbon is CO.

The fractional abundance of HC$_3$N relative to H$_2$ derived from observations is $(4.8\pm0.7)\times10^{-9}$, calculated using the HC$_3$N column density derived by \citet{colzi2024} for C1 and adopting an H$_2$ column density of $N(\mathrm{H}_2)=1.35\times10^{23}$ cm$^{-2}$ from \citet{martin2008}, assuming an uncertainty of 15\% of its value. As shown in Fig.~\ref{fig-abundances}, this abundance cannot be fully reproduced by our models, with the closest value attained at approximately 2$\times$10$^{4}$~yr. As discussed in Sect.~\ref{sec-source}, most of the molecules observed toward G+0.693 are likely formed following a major shock event caused by a cloud–cloud collision. Consequently, part of the missing HC$_3$N may reside on grain surfaces. \citet{dutkowska2025} recently investigated the chemistry of typical CMZ sources, including shocked environments, and demonstrated that the HC$_3$N abundance observed in G+0.693 can be reproduced using a shock model with a cosmic-ray ionization rate of $\sim$10$^{-15}$~s$^{-1}$ (see their Figs.~2 and~4), consistent with values inferred from previous studies. These differing physical conditions could, in principle, influence the predicted isotopic ratios; therefore, they warrant a more detailed investigation in future dedicated works. In any case, sufficient time has elapsed since the shock event (between 10$^{4}$ and 10$^{5}$~yr, see e.g., \citealt{requena-torres2006,sanz-novo2024}) for the gas-phase chemistry to reach a new equilibrium. Therefore, we do not expect significant deviations from the results obtained at comparable timescales, as explained below.

We now define the isotopic fractionation degree, $\varphi$(X), as the ratio between the initial elemental $^{12}$C/$^{13}$C ratio and the predicted $^{12}$C/$^{13}$C ratio for species X. We have tested that this quantity is independent of the initial value adopted in the model. Figure~\ref{fig-fractionationdegree-C3} presents $\varphi$(C$_3$) averaged over its three possible isotopologues corresponding to the position of the $^{13}$C atom. For C1, $\varphi({\mathrm{C_3}})$ ranges between 1.2 and 1.3, while for C2 it spans values from 0.85 up to 3.2. These results are generally consistent throughout most of the chemical evolution with theoretical expectations based on the isotopic exchange reactions $^{13}$C + C$_3 \rightarrow ^{12}$C + C$^{13}$CC/$^{13}$CCC/CC$^{13}$C, which are exothermic by $\Delta E =$ 43~K, 27~K, and 27~K, respectively (e.g., \citealt{colzi2020, loison2020}).
These reactions proceed efficiently as long as C$_3$ remains abundant, that is, at times earlier than $\sim$3$\times$10$^{4}$ yr, causing the fractionation degree $\varphi$(C$_3$) to approach $e^{\Delta E/T_{\rm kin}}$, where $T_{\rm kin}$ is the kinetic temperature of the gas (see Table \ref{table-physprop}). For C1, this yields expected values of 1.36 and 1.21 for the C$^{13}$CC and $^{13}$CCC isotopologues, respectively, while for C2 the corresponding values are 4.19 and 2.45 for C$^{13}$CC and $^{13}$CCC. This is consistent with what is illustrated in Fig.~\ref{fig-fractionationdegree-C3}, which shows that for timescales shorter than $\sim$3$\times$10$^{4}$ yr the predicted $\varphi$(C$_3$) values for C$^{13}$CC and $^{13}$CCC are close to the theoretical values discussed above.

At timescales longer than $\sim$3$\times10^{4}$~yr, we investigate the isotopic fractionation degree of CN and HNC, which also become increasingly important for the formation of HC$_3$N. As shown in Fig.~\ref{fig-fractionationdegree-otherspecies}, both $\varphi$(CN) and $\varphi$(HNC) drop below unity, reaching values as low as $\sim0.6$ in the low-temperature C2 model. At later times, the fractionation degree increases again, approaching values of 1.2--1.3 at $10^{5}$~yr.
This behavior was previously reported by \citet{colzi2020} for models with a high cosmic-ray ionization rate and arises because CO and HCO$^{+}$ act as the main sinks of $^{13}$C during this phase. At longer times, as the abundance of He$^{+}$ produced by cosmic rays increases, atomic carbon and carbon chains are re-formed, and their isotopic fractionation degree returns to higher values comparable to those at earlier times. These effects are more pronounced in the low-temperature model. Even if not affecting the fractionation degree of HC$_3$N, it is worth noting that for the shorter timescales CN and HNC are enhanced in $^{13}$C because of the isotopic exchange reactions (4) and (25) listed in Table 1 by \citet{sipila2023}.
Similar conclusions apply to HC$_5$N, which is mainly formed from C$_3$ and CN, as illustrated in the scheme in Fig.~\ref{fig:hc5n_network} in Appendix \ref{app:formations}.

Considering the fractionation degrees predicted by our models for C1 and C2, the observed values for the two gas components indicate low to intermediate fractionation degree at times earlier than $3\times10^{4}$~yr\footnote{We do not consider later timescales because the HC$_{3}$N abundance decreases to unrealistically low values ($<10^{-11}$), see Fig.~\ref{fig-abundances}.} (1.28 and 3.15 for C1 and C2, respectively). Under these conditions, we have multiplied the observed ratios with the corresponding isotopic fractionation degree and we have found that the initial \cratio\;for the parental material of molecular clouds in the CMZ is constrained to lie in the range 48–82, with the higher values inferred for C2 when only average estimates are considered. However, when the uncertainties on the \cratio\;obtained for C2 component are taken into account, the inferred initial ratio may be as low as 15$\times$3.15=47 (see Table~\ref{table-ratios-hc3n}), making it compatible with a lower-end initial elemental \cratio\;ratio of 48 for both components. Conversely, we cannot exclude significantly higher values; the 1$\sigma$ upper bond of the C2 average (31$\pm$11 individually, or 26$\pm$3 in the ensemble) suggests an initial ratio reaching as high as $\sim$98, potentially overlapping with ratios measured in the local Galaxy (see Fig.~\ref{fig-comparisonCMZ-MW}). As the spectral features from which we inferred the $^{12}$C/$^{13}$C ratio for C2 are partially blended with the main C1 component, the resulting estimates for C2 are subject to higher systematic uncertainty. Consequently, we adopt the value of 48 inferred from the more robust C1 component as the representative initial ratio for molecular clouds in the CMZ for the remainder of the paper.
At later times, the models predict lower levels of isotopic fractionation, corresponding to possible initial \cratio\;values of 35–41 for C1 and 15–32 for C2. However, because the modeled HC$_3$N abundances approach the observed values more closely at earlier times, we do not consider these ratios further in our discussion.

In future studies, molecules less affected by fractionation, such as CO and HCO$^{+}$, could be used to perform this type of analysis, as shown in Fig.~\ref{fig-fractionationdegree-otherspecies}.

\section{Discussion}
\label{discussion}

In this work we have derived a \cratio\;ratio of 48 toward G+0.693 in the CMZ, indicative of the initial value of the material from which this molecular cloud formed. This value is obtained from molecular species whose line emission is optically thin (single isotopologues of HC$_{5}$N and double isotopologues of HC$_{3}$N), and is corrected for isotopic fractionation effects using astrochemical models based on the most up-to-date chemical network including $^{13}$C- and $^{15}$N-isotopologues.
In Sect.~\ref{sec:comparisonothersources} we compare this result with \cratio\;ratio measured both in the Milky Way (CMZ and disc) and in extragalactic sources.
Furthermore, the derived \cratio\;ratio can be used as probe of the origin of the gas, and the elements within it, in the CMZ (Sect.~\ref{sec-nucleosynthesis}).

\subsection{Comparison with other sources}
\label{sec:comparisonothersources}

\subsubsection{Comparison with other CMZ sources}

\begin{figure*}[h!]
\centering
\includegraphics[width=\textwidth, trim=0cm 0cm 0cm 0cm, clip]{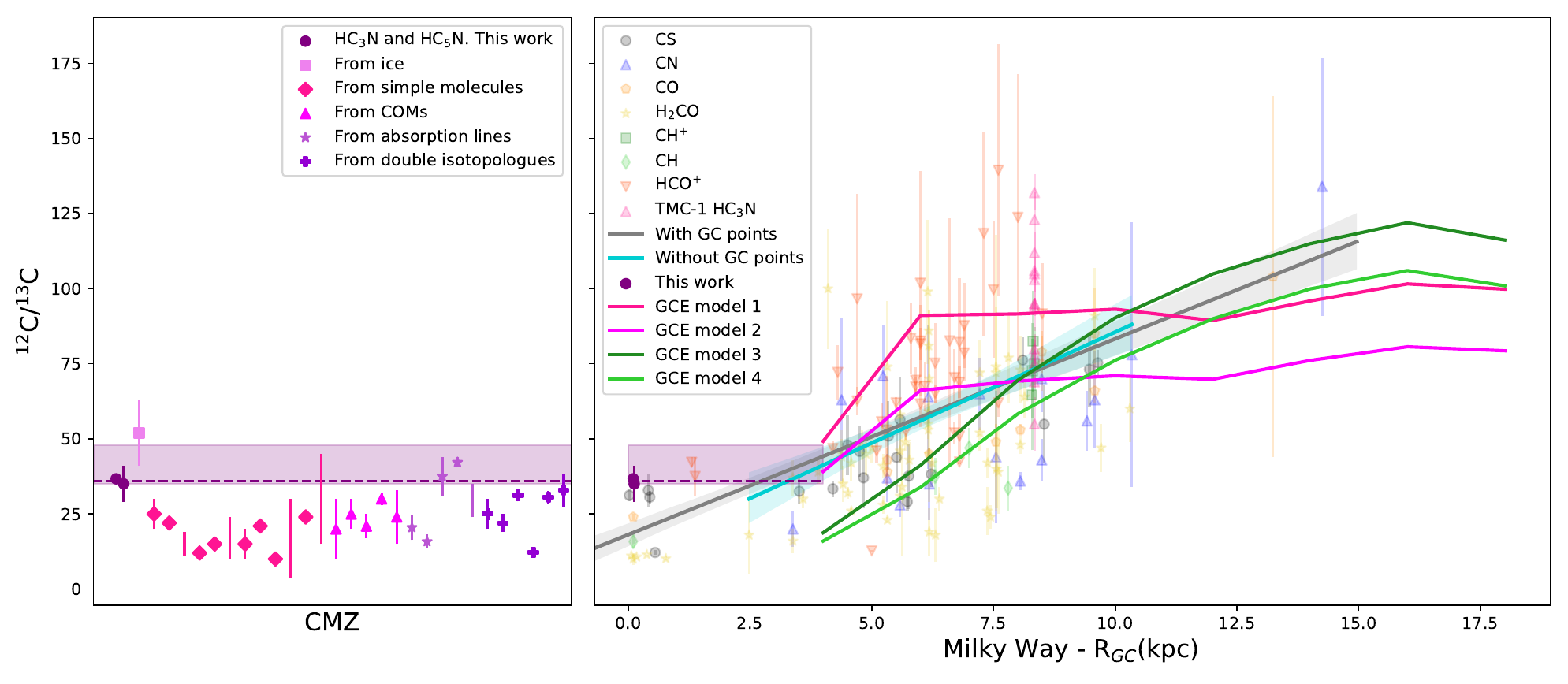}
\caption{$^{12}$C/$^{13}$C ratios measured in the CMZ (left panel) and across the Galactic disc (right panel). In both panels, the purple dashed line indicates the value derived in this work ($\sim$37), while the shaded region represents the range of values obtained when isotopic fractionation effects are taken into account. In the right panel, this line is extended to 4~kpc for comparative purposes. In the left panel, all the sources are within galactocentric distances of 20 pc and 1.3 kpc. Lines without symbols indicate parameter ranges reported in the literature, as opposed to individual measurements shown as symbols. Those measurements are taken from the literature, including values derived from ice observations \citep{boogert2000}; from simple molecules \citep{fomalont1973,wannier1978,gardner1979,goldsmith1981,gardner1982,guesten1985b,langer1990,turner1991,jones2012,riquelme2010,martin2012,yan2019}; from complex organic molecules (COMs) \citep{muller2008,belloche2013,muller2016,margules2016,halfen2017}; from absorption lines \citep{giesen2020,jacob2020,humire2020,luo2024}; and from double-isotopologue methods \citep{humire2020,yan2023}. In the right panel, literature measurements are derived using C$^{18}$O \citep{langer1990,wouterloot1996,keene1998}, CN \citep{savage2002,milam2005}, CS \citep{yan2023}, H$_{2}$CO \citep{henkel1980,henkel1982,henkel1983,henkel1985,yan2019}, CH$^{+}$ \citep{ritchey2011}, CH \citep{jacob2020}, HCO$^{+}$ \citep{luo2024}. Moreover the values derived from different HC$_{3}$N isotopologues toward TMC-1 are taken from \citet{tercero2024}. This panel also includes Galactic chemical evolution (GCE) model predictions from \citet{colzi2022b}. Linear fits to the observational data, including and excluding Galactic Center sources, are shown as grey and cyan solid lines, respectively, with their corresponding shaded confidence regions.}
\label{fig-comparisonCMZ-MW}
\end{figure*}

All measured \cratio\ ratios toward the CMZ are summarized in the left panel of Fig.~\ref{fig-comparisonCMZ-MW}. In the CMZ of the Milky Way, the $^{12}$C/$^{13}$C isotopic ratio has been extensively investigated over the past five decades through radioastronomical observations. Measurements toward prominent molecular clouds such as Sgr~B2 and Sgr~A yield values ranging from $\sim$3 to 30 for relatively simple molecules (e.g., HC$_3$N, H$_2$CO, OCS, HCO$^+$, HCN, HNC, C$_3$, and CH), based on both absorption and emission lines (e.g., \citealt{turner1991,giesen2020}). Higher ratios, typically $\sim$20--32, have been derived from COMs, including CH$_3$OH, CH$_3$CN, CH$_3$CH$_2$OH, C$_2$H$_3$CN, and C$_2$H$_5$CN. Most of these determinations rely on direct comparisons between the line intensities of a main isotopologue and its $^{13}$C-substituted counterpart. However, this approach is strongly affected by optical depth effects and therefore generally provides lower limits to the true isotopic ratio. Notably, values derived from COMs tend to lie at the upper end of the observed range, likely because their transitions are typically more optically thin than those of simpler species (e.g., \citealt{halfen2017}), or because they trace different gas components, such as warmer and denser gas located closer to the protostar.

More reliable measurements based on optically thin tracers and double isotopologue methods have also been reported. For example, \citet{humire2020} and \citet{yan2023} derived the carbon isotopic ratio using C$^{34}$S/$^{13}$C$^{34}$S toward the +50~km~s$^{-1}$ cloud and the Sgr~B2 line of sight, obtaining values between 22--35. These results are consistent with the higher end of previously reported values (see left panel of Fig.~\ref{fig-comparisonCMZ-MW}). In addition, \citet{martin2012} derived ratios of 15--45 toward the circumnuclear disc around Sgr~A*, using CN and correcting for line opacity effects through its hyperfine structure. 
\citet{jones2012} found that integrated intensity ratios between $^{12}$C- and $^{13}$C-bearing isotopologues of HCN, HCO$^{+}$, and HNC are typically significantly lower than 24, confirming that emission from the main isotopologues is generally optically thick in the CMZ. However, they also identified regions at the most redshifted velocities ($\sim$100~km~s$^{-1}$) where the \cratio\;ratio derived from HCO$^{+}$ reaches values as high as 45. Consistently, \citet{riquelme2010} reported \cratio\ ratios $>$40--70 in the external disc–halo interface of the CMZ, suggesting the presence of infalling less processed, more pristine material from a nucleosynthesis perspective.
 
In this work, we derive a ratio of $\sim$37 using HC$_5$N and the double isotopologue of HC$_3$N, and up to 48 when correcting for fractionation effects. Moreover, using the column densities reported by \citet{colzi2022a}, considering HC$^{15}$N and H$^{13}$C$^{15}$N, we derive a \cratio\ ratio of $\sim$36 (not corrected for fractionation) toward the C1 component of G+0.693. More recently, \citet{luo2024} derived \cratio\;ratios higher than 35 from absorption-line observations toward Galactic center region lines of sight ($R_{\rm GC} \sim 1$~kpc).
Finally, \citet{boogert2000} used ISO-SWS ice observations toward the Galactic center source GC3 and obtained a CO$_2$/$^{13}$CO$_2$ ratio of $52\pm11$. These results further support the presence of a \cratio\;ratio of at least $\sim$35 toward the CMZ and highlight the importance of employing optically thin tracers, as well as considering fractionation effects, for accurate determinations. 

\subsubsection{Comparison with Milky Way sources}
\label{sec:MW}

The right panel of Fig.~\ref{fig-comparisonCMZ-MW} compiles the most relevant \cratio\;measurements obtained in molecular clouds across the Galactic disc and in the CMZ.
For the comparison with Galactic sources, we revised the adopted galactocentric distances to ensure consistency with the most recent determinations. For H$_2$CO, CS, CO, and CN, we adopted the updated distances reported by \citet{yan2019,yan2023} , which use trigonometric parallaxes when available and revised kinematic distances otherwise. The only exception is WB89-437, for which \citet{wouterloot1996} derived a $^{12}$CO/$^{13}$CO ratio of $104 \pm 60$; for this source we adopted the trigonometric parallax distance of 13.24 kpc determined by \citet{hachisuka2015}. For CH$^+$ and CH, we used the galactocentric distances reported by \citet{ritchey2011} and \citet{jacob2020}, respectively.
The majority of the sources in the disc lie at Galactocentric distances $R_{\rm GC}$ between $\sim$2.5 and 11~kpc. The data include observations obtained with different molecular tracers; the molecules used and the corresponding references are listed in the figure and in Sect.~\ref{sec:comparisonothersources}. On average, the \cratio\;increases with Galactocentric distance. This behavior is also consistent with the most recent Galactic chemical evolution models \citep{colzi2022b}.

A linear regression fit including all the measurements yields the following 
Galactocentric gradient:
\begin{equation}
^{12}\mathrm{C}/^{13}\mathrm{C} = (18 \pm 3) + (6.5 \pm 0.5)\,R_{\mathrm{GC}},
\end{equation}
when the CMZ data points are included. Excluding the Galactic Center and $R_{\rm GC}>$12 kpc sources results instead in the relation:
\begin{equation}
^{12}\mathrm{C}/^{13}\mathrm{C} = (12 \pm 7) + (7.4 \pm 1.1)\,R_{\mathrm{GC}}.
\end{equation}
These trends are consistent with the observational trend obtained in previous works (see references in caption of Fig.~\ref{fig-comparisonCMZ-MW}) and with the predictions of Galactic chemical evolution models for the Milky Way (e.g., \citealt{romano2019,romano2022,colzi2022b}, see also right panel of Fig.~\ref{fig-comparisonCMZ-MW}). 

Using the two reported linear relations for the Galactic gradient of the 
$^{12}\mathrm{C}/^{13}\mathrm{C}$ ratio, we estimate the Galactocentric 
distances at which the observed values in this work, between 37 and 48, are expected. For the first relation taking into account CMZ sources this interval corresponds to $R_{\mathrm{GC}} \approx 2.9{-}4.6$ kpc when 
using the central values of the coefficients. Similarly, adopting the second relation, excluding CMZ and outer Galaxy sources, the same range of isotopic ratios is obtained at 
$R_{\mathrm{GC}} \approx 3.4{-}4.9$ kpc. Therefore, both gradients consistently 
indicate that $^{12}\mathrm{C}/^{13}\mathrm{C}$ ratios between 37 and 48 are 
most likely found at Galactocentric distances of approximately 
$R_{\mathrm{GC}} \sim 3{-}5$ kpc. When accounting for the uncertainties in the 
slopes and intercepts, this range broadens to roughly 
$R_{\mathrm{GC}} \sim 2.1{-}6.8$ kpc. It should also be noted that the values obtained by \citet{luo2024} at $\sim 1$\,kpc along the line of sight toward the CMZ are consistent with the values derived in our work within 100\,pc of the Galactic Center.

Moreover, in the Galactic disc the values obtained at a given $R_{\rm GC}$ the values obtained are not consistent among them. In fact, they are obtained using observations toward sources with different physical conditions and different molecules, pointing out also in this case the importance of taking into account isotopic fractionation, through astrochemical models, to properly retrieve the initial isotopic ratios of those regions of the Galaxy. For example, locally at 8.2 kpc, different \cratio\;ratios have been evaluated toward the cyanopolyyne peak in TMC-1. \citet{tercero2024} reported that most of the $^{13}$C resides in HCC$^{13}$CN rather than in H$^{13}$CCCN or HC$^{13}$CCN. In that region, the cosmic-ray ionization rate is lower than in the CMZ ($\sim$1.3$\times$10$^{-17}$~s$^{-1}$), making the reaction C$_2$H$_2$ + CN one of the most efficient formation routes for HC$_3$N. Given the low kinetic temperature of TMC-1 (10~K), CN is likely fractionated \citep{colzi2020}, and its low \cratio\;can propagate into HC$_3$N via the aforementioned reaction, which proceeds efficiently only in the forward direction at such low temperatures. The differences in the observed ratios toward TMC-1 indicate that this reaction does not involve carbon exchange during formation. Consequently, \cratio\;values obtained both from the single and double isotopologues when including H$^{13}$CCCN and HC$^{13}$CCN, which exhibit \cratio\;values within the range 78--105 (see right panel of Fig.~\ref{fig-comparisonCMZ-MW}), likely better represent the intrinsic isotopic ratio of the local TMC-1 molecular cloud. In contrast, our results toward G+0.693 show that the \cratio\;values are consistent within 1$\sigma$ among the different $^{13}$C isotopologues of HC$_3$N in both C1 and C2, in line with on average higher $T_{\rm kin}$. However, as mentioned above, proper chemical models are needed to interpret the results (see Sect.~\ref{sec:astrochemistry}).

\subsubsection{Comparison with external galaxies}
\label{sec:comparisonextragal}

\begin{figure*}[h!]
\centering
\includegraphics[width=\textwidth, trim=0cm 0cm 0cm 0cm, clip]{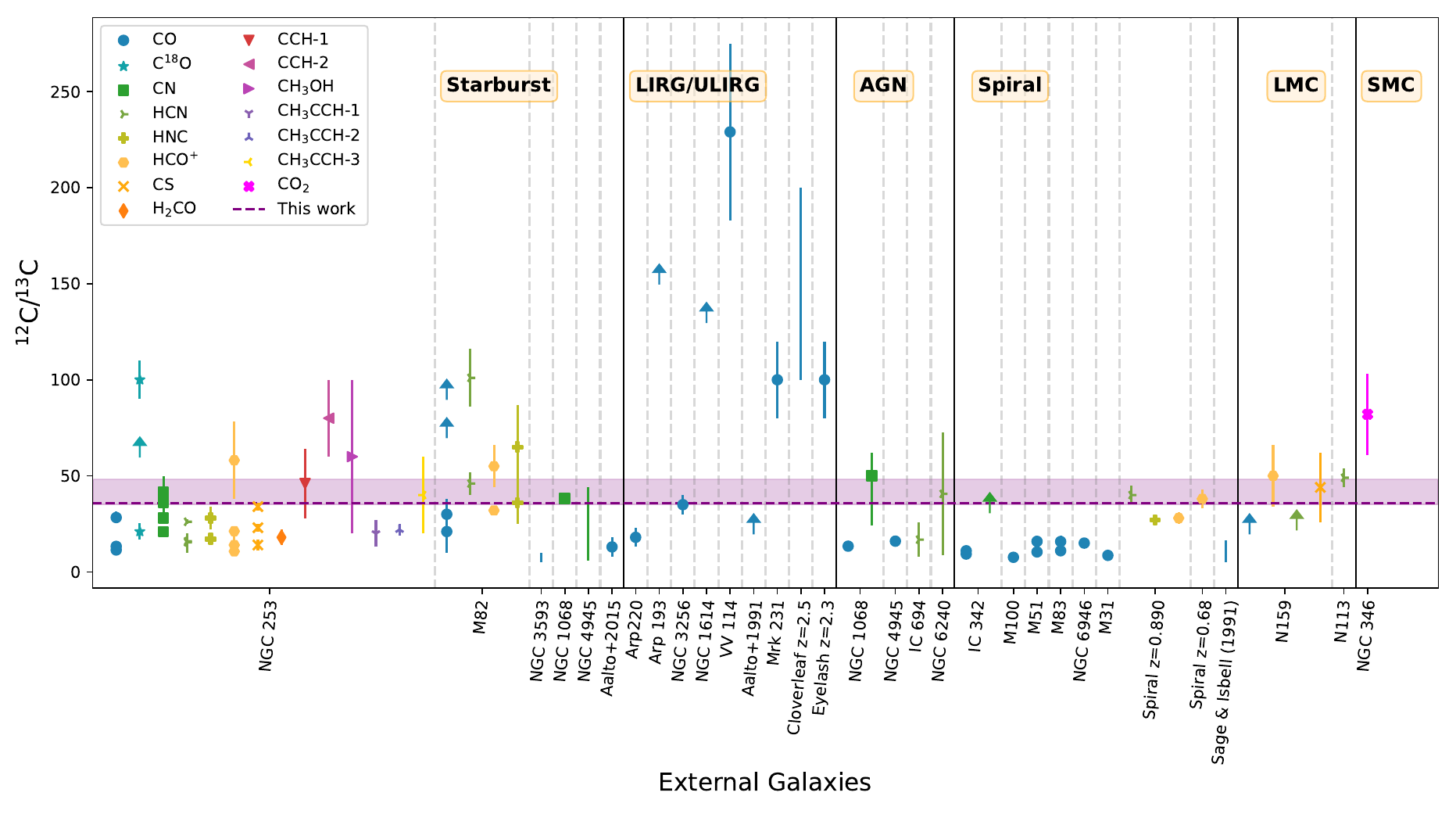}
\caption{$^{12}$C/$^{13}$C ratios measured in external galaxies. The purple dashed line indicates the value derived in this work ($\sim$37), while the shaded region represents the range of values obtained when isotopic fractionation effects are taken into account. Lines without symbols indicate parameter ranges reported in the literature, as opposed to individual measurements shown as symbols. Values are taken from the literature, including values from starburst and spiral galaxies (\citealt{wiklind1990,young1986,encrenaz1979,stark1984,loiseau1990,sage1991,wiklind1992,henkel1993,aalto1995,kikumoto1998,muller2006,martin2010,henkel2014,aladro2015,wallstrom2016,martin2019b,tang2019,martin2021,li2022}), from LIRGs/ULIRGs (\citealt{aalto1991a,aalto1991b,sliwa2013,danielson2013,sliwa2014,henkel2014,papadopoulos2014,spilker2014}), from AGN \citep{whiteoak1990,solomon1990,jiang2011,aladro2013,tang2019}, from the LMC \citep{boreiko1991,johansson1994,wang2009}, and the SMC \citep{demarchi2025}. Numbers in the legend correspond to the position of the $^{13}$C in the molecule with respect to the C atoms.} 
\label{fig-externalgalaxies}
\end{figure*}

In external galaxies, the \cratio\;ratio spans nearly an order of magnitude (Fig.~\ref{fig-externalgalaxies}). Values exceeding 100 are reported in luminous and ultraluminous infrared galaxies (LIRGs/ULIRGs), whereas starburst, AGN, and normal spiral galaxies typically exhibit lower ratios. Such lower values are generally associated with systems that are not undergoing large-scale mergers. Early extragalactic measurements were primarily based on $^{12}$CO and $^{13}$CO. However, the main isotopologue is usually severely optically thick ($\tau > 3$; e.g., \citealt{young1986,wiklind1990,krugel1990}), often yielding \cratio\;ratios below 15. Thus, we will not discuss the ratios derived from CO further in this paper. 
Subsequent studies employing alternative tracers reported higher and likely more reliable values. \citet{aalto1991a,aalto1991b} derived a ratio of 35$\pm$5 toward the center of the LIRG NGC~3256 and $>$20 toward four merging ULIRGs. They proposed that the elevated intensity ratios reflect a reduction in the mean optical depth of molecular clouds, possibly due to tidal disruption or starburst activity triggered by the merger process.
A more robust determination was presented by \citet{henkel1993}, who reported \cratio\;$>$30 within the central 500~pc of the starburst galaxy NGC~253 using CN, CS, and HNC. Similarly, \citet{mauersberger1993} suggested that enhanced central \cratio\;ratios are most plausibly explained by the inflow of less chemically processed gas toward the nuclear region. Such inflows may be driven by galaxy interactions, mergers, or by the dynamical effects of a stellar bar (see Sect.~\ref{sec-nucleosynthesis}). Consistent results were later obtained by \citet{tang2019}, who derived values of $\sim$40 in three nearby starburst galaxies using CN. They also proposed that gas inflow from the outer disc into the circumnuclear region (inner kpc), possibly mediated by the halo and bar, could explain the observed enhancements. 

More recently, spatially resolved observations of the starburst galaxy M82 have revealed a gradient in the \cratio\;ratio along its major axis, with values of $\sim$40 in the inner regions (very similar to the values obtained in our work) increasing to $\sim$50--100 at 0.45 kpc from the center (\citealt{li2022}). This trend may reflect reduced optical depth effects outside the central regions, as well as genuine nucleosynthesis variations, as discussed in the following section for the CMZ.
The ALMA ALCHEMI large program has provided a complete astrochemical census of the inner 500~pc of the nearby starburst galaxy NGC~253. Using ACA data, \citet{martin2021} derived \cratio\;ratios from a wide range of molecular species, including CO, C$^{18}$O, HCO$^{+}$, HCN, HNC, CN, CS, H$_2$CO, CCH, and HC$_3$N, among others. They found a huge spread of values, demonstrating that the inferred isotopic ratios depend both on the molecular tracer and on the spatial resolution of the observations (Fig.~\ref{fig-externalgalaxies}; see also \citealt{meier2015,martin2019b}). This result underscores the necessity of accounting for line optical depth effects, varying molecular spatial distributions, as well as possible fractionation effects when deriving isotopic ratios also in extragalactic environments.

The values discussed above generally correspond to measurements averaged over the inner central kiloparsec of these galaxies, making them directly comparable to those obtained in the CMZ of the Milky Way (Fig.~\ref{fig-externalgalaxies}). The \cratio\;ratio derived in this work is consistent with the values reported for starburst, AGN, and spiral galaxies (excluding ratios derived from the optically thick CO lines), suggesting that similar gas processing and chemical enrichment mechanisms may operate across these environments, as discussed in Sect~\ref{sec-nucleosynthesis}. However, this comparison should be taken with cautions since the values from extragalatic sources taken from the literature could be both affected by line opacity and isotopic fractionation effects (e.g. \citealt{viti2020}).

Measurements toward the nearby dwarf galaxies, the Large Magellanic Cloud (LMC) and the Small Magellanic Cloud (SMC), show values $>20$ \citep{boreiko1991,johansson1994,wang2009}, reaching up to 82 in more recent  measurements based on CO$_2$ ice observations \citep{demarchi2025}. Possible connections of the CMZ with these systems are also discussed in the next section, together with the implications for chemical nucleosynthesis evolution.

\subsection{Implications of the carbon isotopic ratios for Galactic center evolution and stellar nucleosynthesis}
\label{sec-nucleosynthesis}

\subsubsection{Carbon isotopic ratios as probes of gas accretion in the Galactic center}
\label{sec:gasaccretion}

In Sect.~\ref{sec:MW} we have discussed that comparison with the \cratio\;ratios derived in the Milky Way disc toward massive star-forming regions shows that the values obtained in this work toward the CMZ are similar to those at galactocentric distances of $\sim$3--5 kpc (see right panel of Fig.~\ref{fig-comparisonCMZ-MW}). 
This supports a scenario in which the CMZ is replenished by gas with isotopic ratios characteristic of larger Galactocentric distances. 
This process can be explained by the presence of bars as one of the main mechanisms driving gas inflow from the Galactic disc toward the CMZ where the dense molecular gas accumulates (e.g., \citealt{gerhard2011,sormani2019,tress2020,henshaw2023}). 
It can be noted that a part of this flow can also come from the halo, a roughly spherical region surrounding the Galaxy that contains hot ionized gas, dark matter, and some old stars, where the gas is less reprocessed. In fact, there are some evidences that unmixed stars within the halo present \cratio\;ratios higher than 40 (e.g., \citealt{molaro2023, rizzuti2025}). This inflow can occur through the infall of cooler gas clouds, large-scale accretion along filaments, or galactic fountains driven by supernova explosions in the galactic disc (e.g., \citealt{henley2010,riquelme2010,bish2019}). If the CMZ is fed by gas coming from a few kpc out in the disc, its isotopic ratios can reflect those larger $R_{\rm GC}$ values rather than purely local nucleosynthesis. 

Another possible mechanism is the past interaction of the Milky Way with dwarf galaxies. The Milky Way has undergone several merger events with dwarf galaxies during its evolution. A well-known example is the ongoing accretion of the Sagittarius dwarf spheroidal galaxy, whose tidal disruption has produced the Sagittarius stellar stream \citep{ibata1994,majewski2003}. In addition, analyses of stellar kinematics and chemical abundances from \textit{Gaia} data have revealed evidence for a major ancient merger with a dwarf galaxy known as Gaia--Enceladus, which contributed significantly to the formation of the Galactic stellar halo \citep{Helmi2018,myeong2018,berni2026}. Such accretion events illustrate how external material can be incorporated into the Milky Way and potentially influence the chemical properties of its gas reservoir. In fact, dwarf galaxies typically exhibit lower metallicities and contain less chemically processed material than the Milky Way, which can modify isotopic ratios once this gas is accreted into the inner Galaxy. Recent work suggests that star formation in the inner Milky Way has occurred in several episodic bursts rather than continuously. Using Gaia color–magnitude diagram fitting, \citet{ruizlara2026} showed that super-metal-rich stars in the solar neighbourhood, likely formed in the inner Galaxy and later migrated outward, trace multiple star formation episodes over the past $\sim13$ Gyr, possibly triggered by satellite interactions and gas inflow events (see also \citealt{palla2024}).
After mixing within the Galactic halo or disc and subsequently migrating inward, such gas may increase isotopic ratios in the Galactic center, as observed here for carbon. This scenario is supported by the similarities between the isotopic ratios measured in the Large and Small Magellanic Clouds and those derived in this work (Sect.~\ref{sec:comparisonextragal} and Fig.~\ref{fig-externalgalaxies}).

\subsubsection{Some insights from stars and chemical evolution models of the CMZ}

From the perspective of stellar nucleosynthesis, $^{12}$C is predominantly synthesized via the triple-$\alpha$ process during the giant and supergiant phases of low-, intermediate-, and massive stars. In contrast, $^{13}$C is primarily a secondary isotope produced through the CNO cycle and in nova outbursts. The enrichment of $^{13}$C is largely dominated by long-lived asymptotic giant branch (AGB) stars and therefore occurs on longer timescales than the enrichment of $^{12}$C. This difference in production channels naturally leads to a decrease in the carbon isotope ratio with increasing metallicity and decreasing $R_{\rm GC}$ across the Galactic disc. These trends are consistent with the predictions of Galactic chemical evolution models for the Milky Way disc (e.g., \citealt{romano2019,romano2022,colzi2022b}, see also right panel of Fig.~\ref{fig-comparisonCMZ-MW}). In this context, Galactic chemical evolution models provide an essential framework for interpreting isotopic ratios and their variation across different galactic environments. In particular, models that focus on the central regions of galaxies, including the CMZ of the Milky Way, aim to describe how processes such as gas inflow, star formation, stellar feedback, and mixing shape the chemical enrichment history of the nuclear regions (e.g., \citealt{spitoni2026}). In fact, all of these are processes that could explain the high \cratio\;ratio of 48, typical of less processed material, inferred for the CMZ from this work.

Chemical evolution models focusing on the central zones of our Galaxy can only be constrained by a few measurements toward stars. For example, \citet{kovtyukh2019,kovtyukh2022} measured from Cepheids [Fe/H]$<$0.4 dex in the CMZ with a decrease in the inner 2 kpcs from 0.4 dex to Solar 0 dex value. In general, stars in the Nuclear Stellar disc (NSD), the stellar counterpart of the CMZ, present Solar or slightly higher metallicity (see also \citealt{najarro2009,ryde2015,schultheis2019,thorsbro2020, matsunaga2023}). These measurements further support the idea of a possible inflow of less reprocessed and low metallicity gas in the central Galaxy. Going at smaller scales \citet{chen2023} measured [Fe/H]$\sim$0.5 dex or less in the inner 1.5 pc in the nuclear star cluster (NSC) in the vicinity of the supermassive black hole. In fact, the sub-solar [Fe/H] stars have spatial anisotropy indicating a recent star cluster infall event, as well as they have different kinetics from the super solar-ones (\citealt{feldmeier-krause2020,do2020}). They suggested that this is consistent with an infall of a globular cluster or of a dwarf galaxy.

From the point of view of models, \citet{grieco2015} studied the chemical evolution of the NSD and compared with $\alpha$-elements and Fe abundances of M giant stars by \citet{ryde2015}. They suggest that the Galactic Center experienced a rapid early formation phase with a high star formation efficiency ($\sim25\,\mathrm{Gyr^{-1}}$) and an Initial Mass Function (IMF) richer in massive stars than the standard solar-neighborhood IMF (e.g., \citealt{kroupa1993}), with the \citealt{chabrier2003} IMF providing a better match to observations). The present-day star formation rate could be explained by a secondary burst triggered by a modest gas infall episode 0.5 Gyr ago, likely associated with bar-driven accretion, without significantly affecting the abundance patterns or the metallicity distribution function.

A more recent study by \citet{friske2025} further investigated the chemical evolution of nuclear stellar discs using analytical multi-zone models tailored to the central regions of barred galaxies. Their work explores how sustained gas inflow toward the Galactic center can shape the metallicity distribution and abundance patterns observed in NSDs. They find that continuous or recurrent gas accretion from the outer parts of the galaxy into the nuclear region, likely driven by the Galactic bar, can sustain star formation over extended periods while progressively enriching the interstellar medium. In this scenario, the NSDs are not formed in a single early episode but rather through prolonged star formation fueled by inward gas transport, yielding the metallicities and complex abundance trends usually observed in nuclear stellar populations.

Recent constraints on the chemical composition of the NSD were provided by \citet{thorsbro2020}, who analised high-resolution near-infrared spectra of M giant stars in the nuclear region. They found a predominantly metal-rich population, with metallicities up to $\mathrm{[Fe/H]} \sim +0.5$, and $\alpha$-element abundances indicating a complex enrichment history. By comparing their measurements with chemical evolution models of the Galactic Center by \citet{grieco2015}, they argued that the observed abundance patterns are consistent with rapid early enrichment (mini-starburst activity) followed by a more recent starburst $\sim 1$\,Gyr ago, involving accretion of primordial or slightly enriched gas. To trigger this starburst episode, the gas must be accreted either from other regions of the Galaxy or from extragalactic origin. This mini-starburst scenario is further supported by our findings, as the isotopic ratios we derive toward G+0.693 closely resemble those measured in starburst galaxies.

More recently, \citet{spitoni2026} presented a new chemical evolution model for the NSD within a Bayesian framework. In this work, the authors explored scenarios in which the NSD forms primarily from gas transported to the Galactic centre through bar-driven inflows from the inner disc. By comparing model predictions with the observed metallicity distribution function and $\alpha$-element abundance ratios, they showed that a simple inner-disc origin struggles to reproduce the metal-poor tail of the data. Models including additional dilution from more metal-poor gas, or alternatively adopting a low star formation efficiency combined with mild galactic winds, provide a better agreement with the observations, although the results depend on the level of bulge contamination in the stellar samples.

These results highlight the uncertainties in the chemical evolution of the NSD and the need for additional constraints beyond stellar abundances alone. In this context, measurements of the chemical composition of the interstellar medium, particularly in molecular clouds, provide a complementary probe of the ongoing enrichment processes in the Galactic centre. In particular, isotopic ratios, such as the $^{12}\mathrm{C}/^{13}\mathrm{C}$ studied in this work, are powerful diagnostics of nucleosynthetic enrichment and gas processing, as they trace the relative contributions of massive stars and intermediate-mass stars over time. Constraining these isotopic ratios in molecular clouds therefore provides important independent constraints on the star formation history, gas inflow, and mixing processes that shape the chemical evolution of the NSD.

\section{Conclusions}
\label{conclusions}

We investigated the $^{12}$C/$^{13}$C isotopic ratio toward the CMZ molecular cloud G+0.693$-$0.027 using observations of HC$_3$N, HC$_5$N, and their $^{13}$C isotopologues. Our main findings are summarized as follows:

\begin{enumerate}

\item Using several $^{13}$C isotopologues of HC$_3$N and HC$_5$N, we derived the $^{12}$C/$^{13}$C ratio in G+0.693. For HC$_3$N, the main C1 extended, warm, and more diffuse velocity component ($T_{\rm kin}$=140 K, and $n_{{\rm H}_2}$=2$\times$10$^{4}$ cm$^{-3}$) yields $^{12}$C/$^{13}$C=36.7$\pm$1.02 based on double-$^{13}$C isotopologues. Moreover, the compact, colder, and denser C2 component ($T_{\rm kin}$=30 K, and $n_{{\rm H}_2}$=5$\times$10$^{4}$ cm$^{-3}$) gives $^{12}$C/$^{13}$C=26$\pm$3. For HC$_5$N, the average ratio obtained for C1 is $^{12}$C/$^{13}$C=38.8$\pm$1.5. Ratios derived from the main HC$_3$N isotopologue are strongly affected by line optical depth effects. The most reliable value for C1 comes from double-$^{13}$C isotopologues of HC$_3$N, as well as from single isotopologues of HC$_5$N. The combined average ratio of $37.7\pm1.0$ for C1 indicates that the carbon isotopic ratio in the CMZ is likely higher than previously adopted.

\item We used astrochemical modeling with the most up-to-date chemical models including $^{13}$C- and $^{15}$N-isotopologues, which suggests that the observed ratios correspond to low-to-intermediate isotopic fractionation degree and are consistent with early chemical times ($\lesssim 3\times10^{4}$ yr) after the shock event affecting G+0.693. This provides a constraint on the initial $^{12}\text{C}/^{13}\text{C}$ ratio of the parent material of the CMZ clouds of $\sim$48.

\item The $^{12}$C/$^{13}$C ratio derived in this work ($\sim$37, increasing up to $\sim$48 when correcting for fractionation effects) lies at the upper end of the values reported for the CMZ and is consistent with measurements obtained using optically thin tracers and double–isotopologue methods. When compared with the Galactic gradient of the $^{12}$C/$^{13}$C ratio, these values correspond to those typically found at Galactocentric distances of $\sim$3--5 kpc, suggesting the presence of less chemically processed gas in the central region. Moreover, the derived ratios are comparable to those observed in the nuclear regions of nearby starburst and spiral galaxies, indicating that similar gas inflow and chemical enrichment processes may operate in galactic centers.

\item The high $^{12}$C/$^{13}$C ratios derived in this work support a scenario in which the CMZ is replenished by less chemically processed gas, originating from larger Galactocentric distances, or from the Galactic halo, with a possible contribution from the merging of external dwarf galaxies. From the perspective of stellar nucleosynthesis and chemical evolution models, such elevated ratios are consistent with the presence of recently accreted material that has experienced limited $^{13}$C enrichment from long-lived AGB stars. Our results therefore provide independent constraints on the chemical evolution of the NSD, supporting scenarios in which sustained or episodic gas inflow from the inner Galactic disc fuels star formation and shapes the isotopic composition of the CMZ.

\end{enumerate}

This study emphasizes the importance of combining different species, multiple isotopologues (including double ones), and astrochemical modeling to accurately derive and interpret isotopic ratios and, as a consequence, the chemical evolution of elements in chemically rich environments such as the CMZ. Such isotopic diagnostics are also becoming powerful tools in extragalactic studies, where measurements of molecular isotopic ratios in starburst galaxies can constrain the stellar initial mass function and the star formation history of distant systems (e.g., \citealt{zhang2018}).

\begin{acknowledgements}
We thank the referee for the valuable comments and suggestions that improved the quality of the paper.

We are very grateful to the Yebes 40m and IRAM 30m telescope staff for their precious help during the different observing runs. The 40m radio telescope at Yebes Observatory is operated by the Spanish Geographic Institute (IGN; Ministerio de Transportes, Movilidad y Agenda Urbana). IRAM is supported by INSU/CNRS (France), MPG (Germany) and IGN (Spain).

The project that gave rise to these results received the support of a fellowship from the ”la Caixa” Foundation (ID 100010434). The fellowship code is LCF/BQ/PR25/12110012.
L.C, I.J.-S., V.M.R., and M.S.-N acknowledge support from the grant PID2022-136814NB-I00 funded by the Spanish Ministry of Science, Innovation and Universities/State Agency of Research MICIU/AEI/10.13039/501100011033 and by ERDF, UE. 
O.S. acknowledges the financial support of the Max Planck Society.
I.J.-S. also acknowledges support by ERC grant OPENS, GA No. 101125858, funded by the European Union. Moreover, I.J.-S, L.C. and S.Z. acknowledge support from CSIC through ILINK project SENTINEL (ILINK23017) and Bilateral project SOULMATE (BIJSP25017). V.M.R. also acknowledges the grant CNS2023-144464 funded by MICIU/AEI/10.13039/501100011033 and by “European Union NextGenerationEU/PRTR”.
M.S.-N. also acknowledges funding from the Alexander von Humboldt foundation under a Humboldt Research Fellowship.

We thank Dr. Donatella Romano and Prof. Serena Viti for fruitful discussions that helped to improve the interpretation of our results.
\end{acknowledgements}

%

\bibliographystyle{aa} 
\bibliography{bibliography} 


\begin{appendix}




\onecolumn

\section{Supplementary observed spectra and model fits}
\begin{figure*}[h]
\centering
\includegraphics[width=30pc]{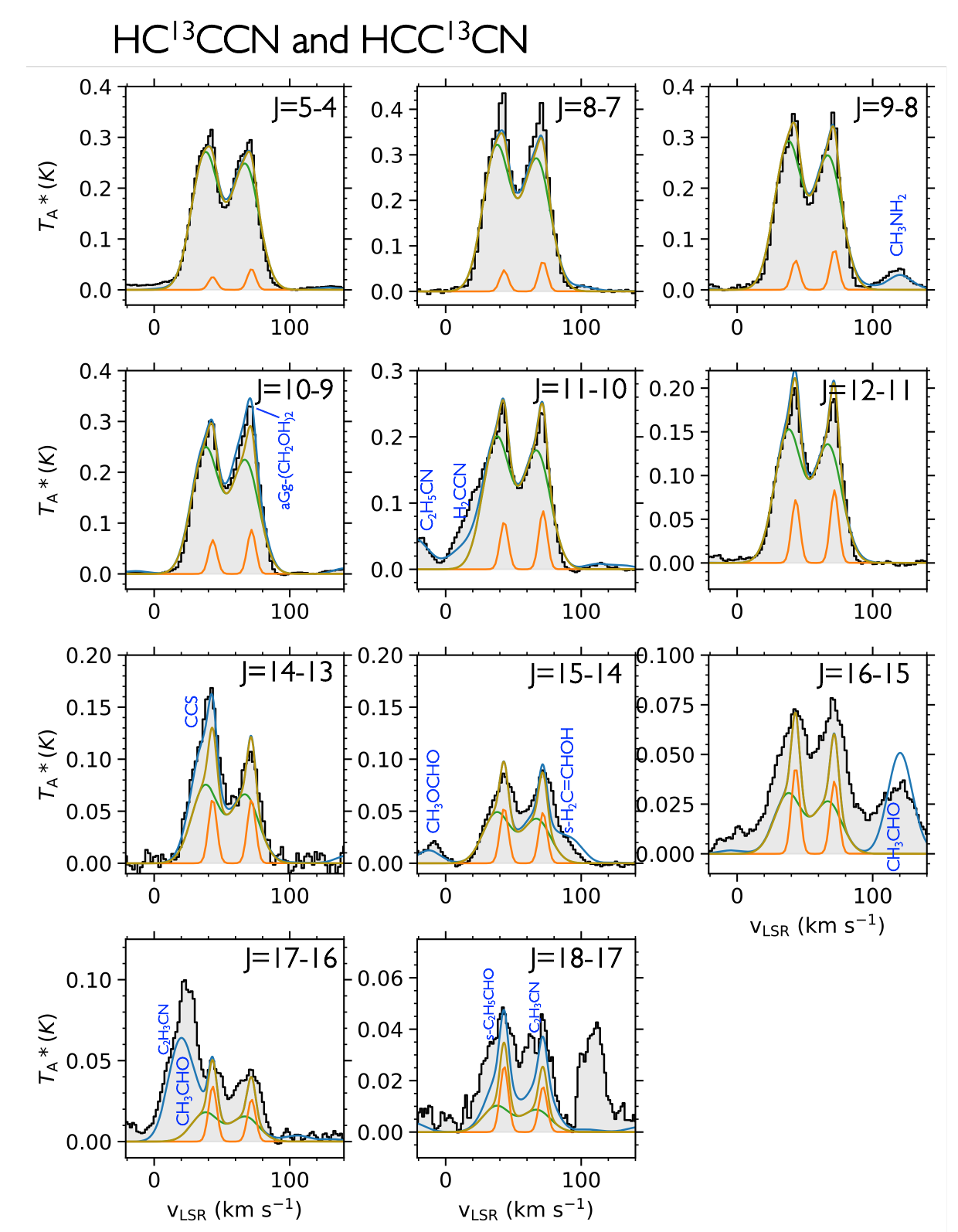}
\caption{Same as Fig.~\ref{fig-h13cccn-fit} but for HC$^{13}$CCN and HCC$^{13}$CN.}
\label{fig-hc13ccn-fit}
\end{figure*}

\begin{figure}[h!]
\centering
\begin{minipage}{0.48\textwidth}
    \centering
    \includegraphics[width=\linewidth]{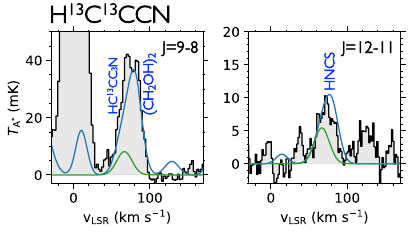}
    \caption{Same as Fig.~\ref{fig-hc13c13cn-fit} but for H$^{13}$C$^{13}$CCN. As explained in Sect.~\ref{sec:analysis-results}, this molecules is tentatively detected and we consider its column density as an upper limit. }
    \label{fig-h13c13ccn-fit}
\end{minipage}\hfill
\begin{minipage}{0.48\textwidth}
    \centering
    \includegraphics[width=\linewidth]{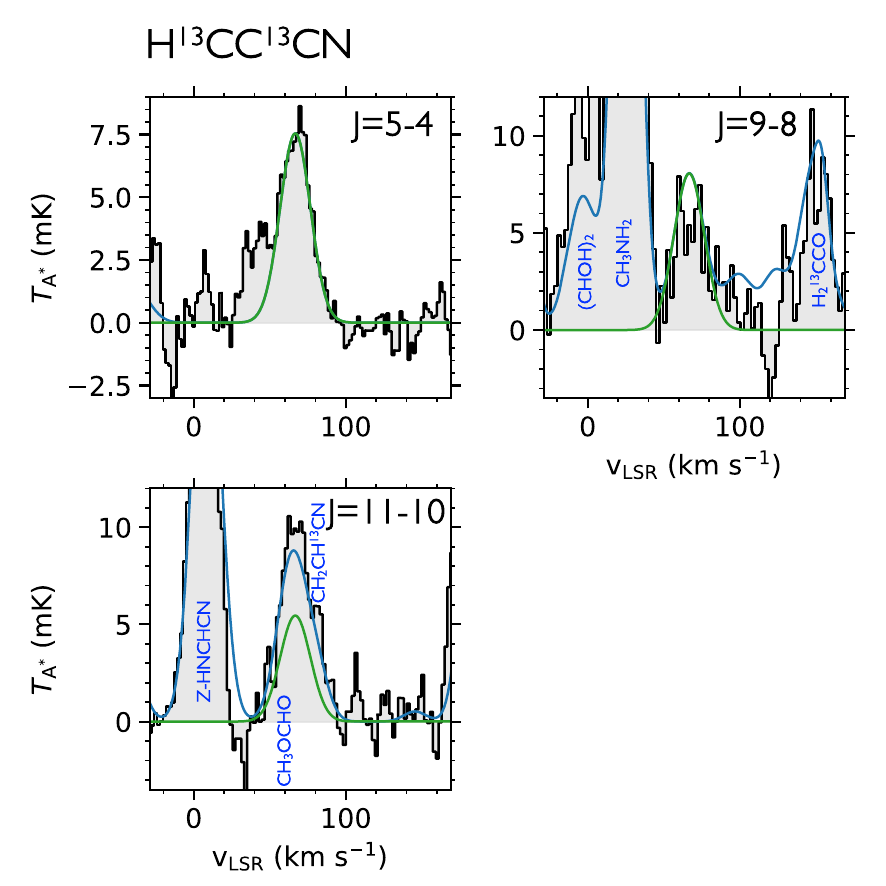}
    \caption{Same as Fig.~\ref{fig-hc13c13cn-fit} but for H$^{13}$CC$^{13}$CN.}
    \label{fig-h13cc13cn-fit}
\end{minipage}
\end{figure}

\begin{figure*}[h!]
\centering
\includegraphics[width=0.85\textwidth]{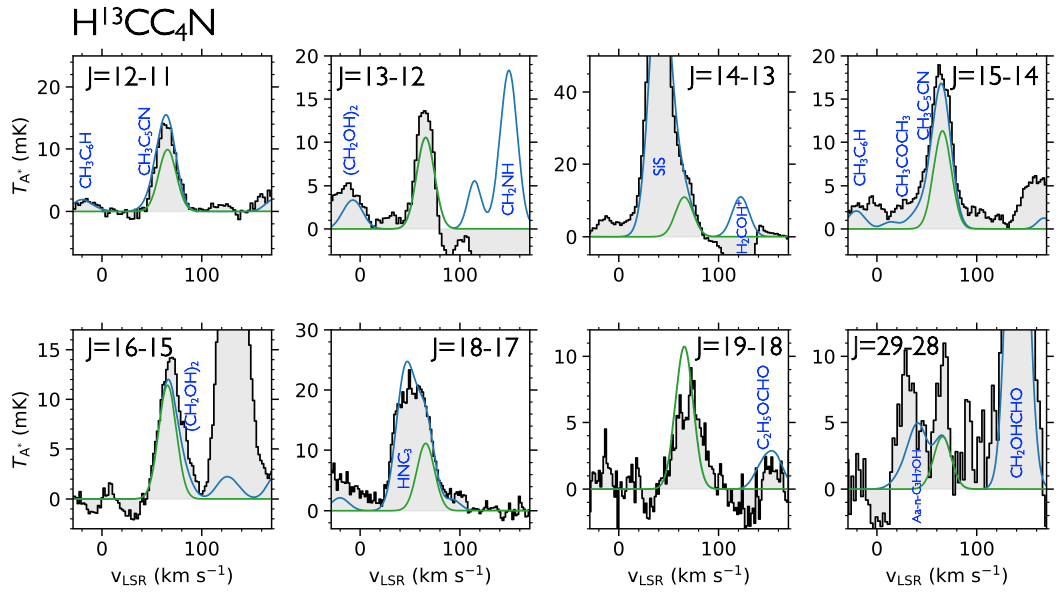}
\caption{Same as Fig.~\ref{fig-hc13c13cn-fit} but for H$^{13}$CC$_{4}$N.} 
\label{fig-h13cc4n-fit}
\end{figure*}

\begin{figure*}[h!]
\centering
\includegraphics[width=0.85\textwidth]{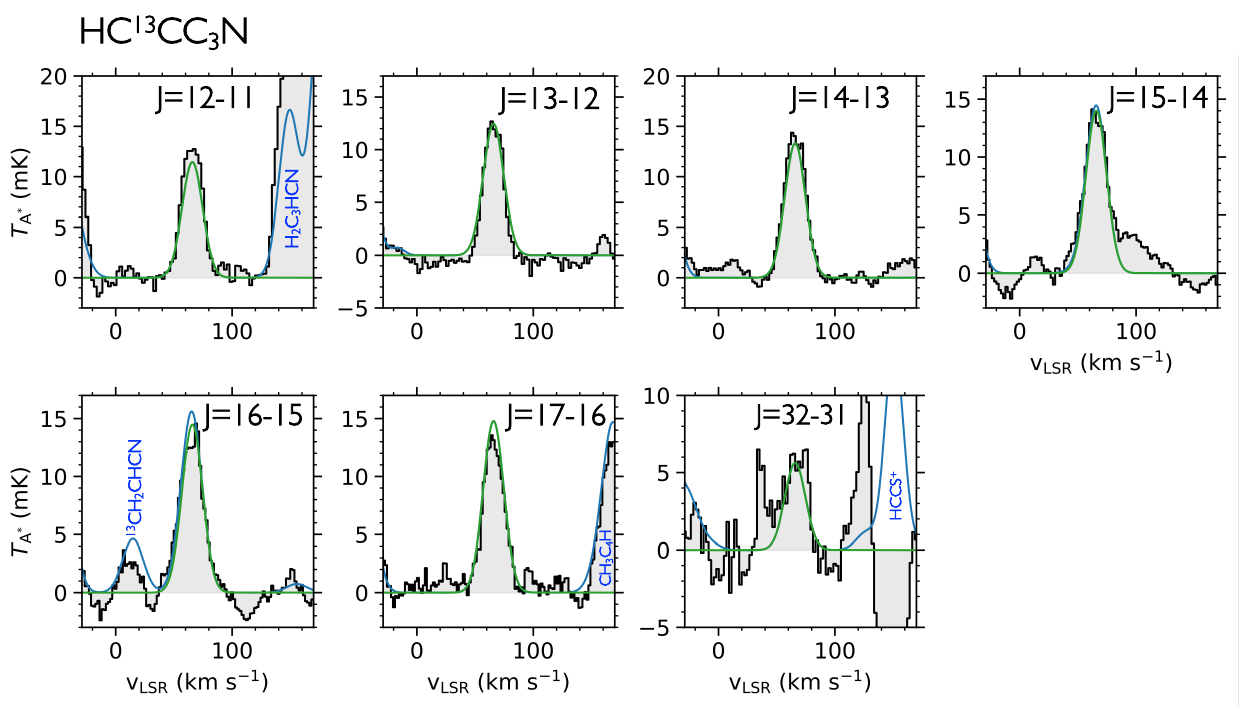}
\caption{Same as Fig.~\ref{fig-hc13c13cn-fit} but for HC$^{13}$CC$_{3}$N.}
\label{fig-hc13cc3n-fit}
\end{figure*}

\begin{figure*}[h!]
\centering
\includegraphics[width=0.85\textwidth]{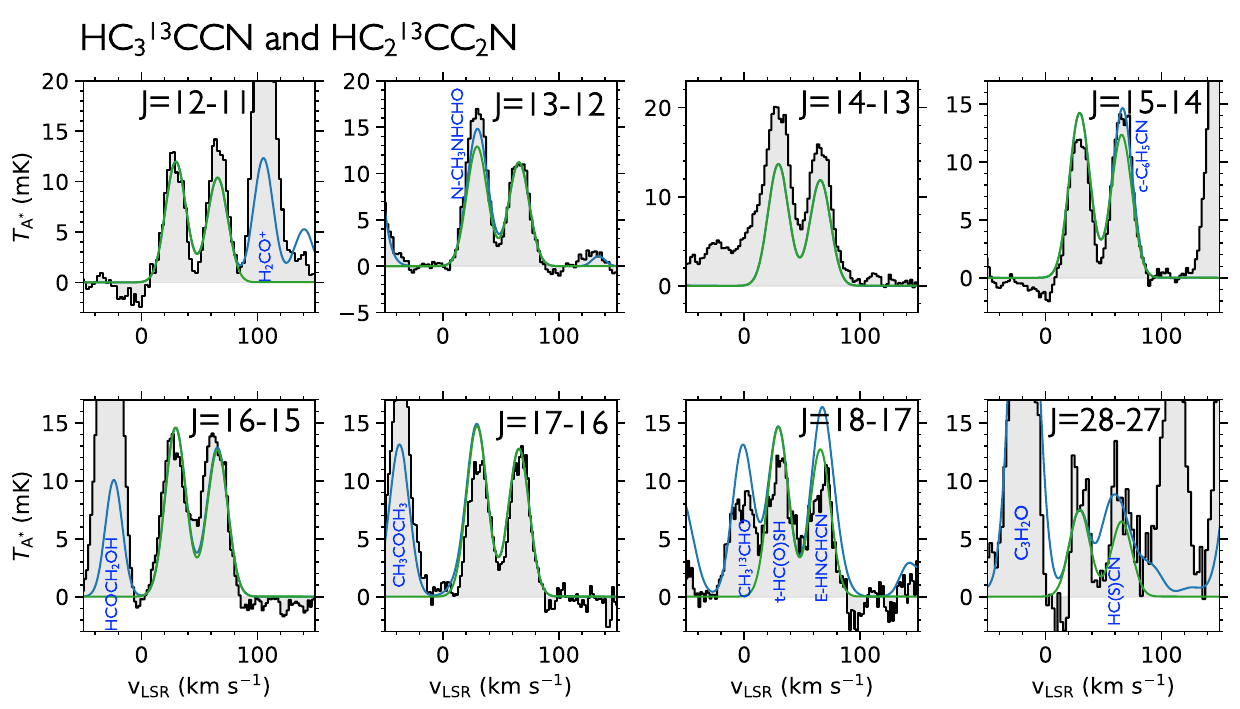}
\caption{Same as Fig.~\ref{fig-hc13c13cn-fit} but for HC$_{3}^{13}$CCN and HC$_{2}^{13}$CC$_{2}$N.}
\label{fig-hc213cc2n-fit}
\end{figure*}

In this appendix we show the observed spectra and the fit to those molecules studied in this work but not shown in the main text: HC$^{13}$CCN, HCC$^{13}$CN, H$^{13}$C$^{13}$CCN, H$^{13}$CC$^{13}$CN, H$^{13}$CC$_{4}$N, HC$^{13}$CC$_{3}$N, HC$_{3}^{13}$CCN, and HC$_{2}^{13}$CC$_{2}$N (Figs.~\ref{fig-hc13ccn-fit}--\ref{fig-hc213cc2n-fit}).

\clearpage
\section{Observed transitions of HC$_5$N in G+0.693}
\label{app-hc5n}
\begin{figure*}[h!]
\centering
\includegraphics[width=0.85\textwidth]{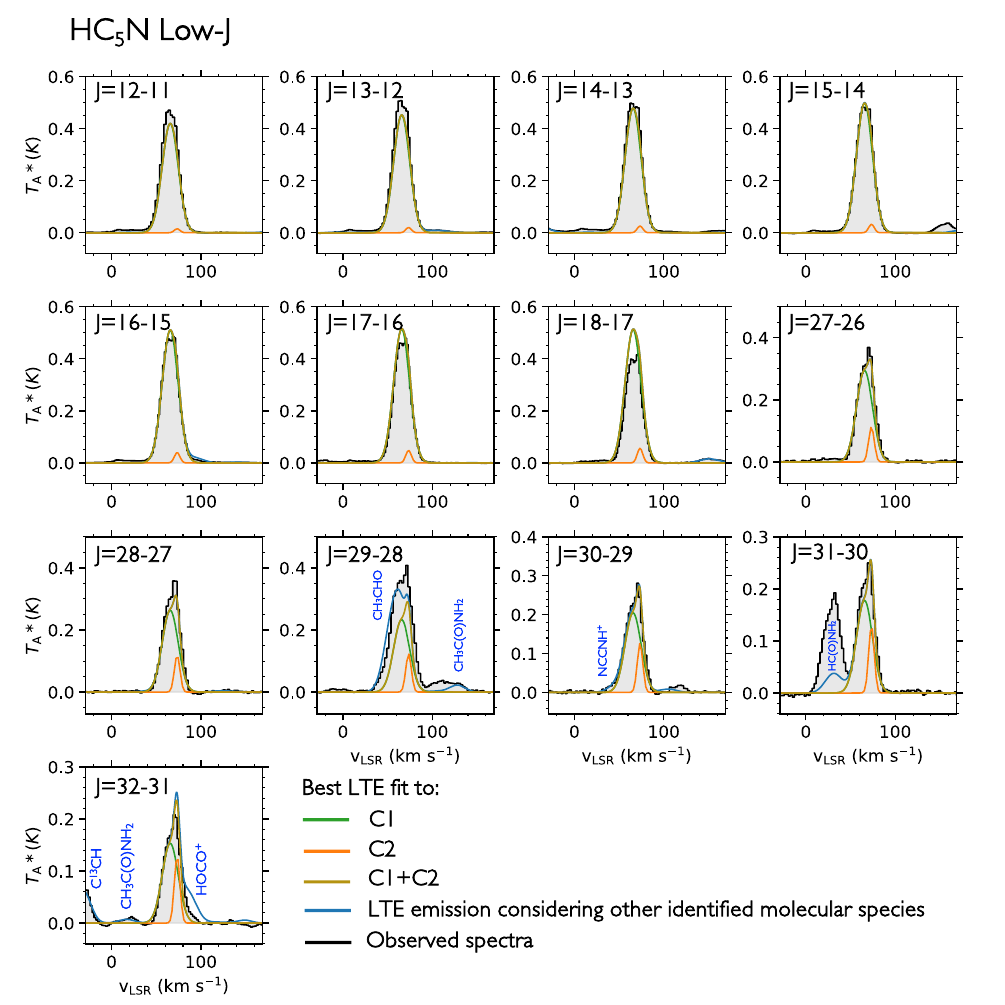}
\caption{HC$_{5}$N observed "low-$J$" transitions used for the LTE analysis (black histrogram).  For each panel, the corresponding transition is indicated in the upper right corner and the telescope information is given if the same transition is observed with more than one instrument. The green and orange solid lines are the best LTE fit to the C1 and C2 components, respectively. The dark gold line is the sum of the two components. The blue line indicates the total modeled line emission, including also the contribution of all molecular species previously identified in the survey (e.g., \citealt{zeng2018,rodriguez-almeida2021a,rodriguez-almeida2021b,rivilla2021a,sanz-novo2023}).}
\label{fig-hc5n-low}
\end{figure*}

\begin{figure*}
\centering
\includegraphics[width=0.85\textwidth]{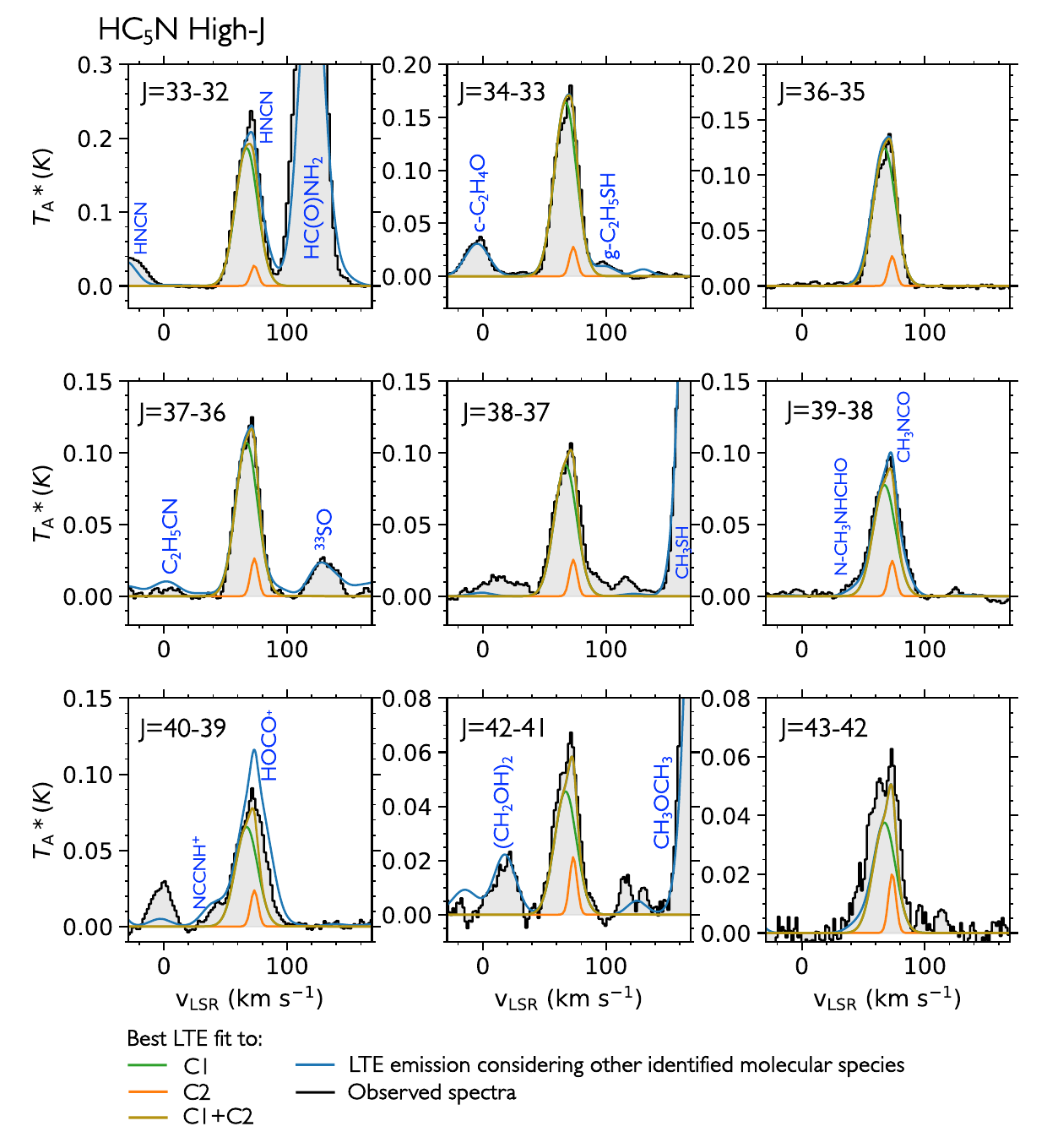}
\caption{Same as Fig.~\ref{fig-hc5n-low} but for "high-$J$" transitions.}
\label{fig-hc5n-high}
\end{figure*}

In this appendix, we present the transitions of HC$_{5}$N detected toward G+0.693 and analised in Sect.~\ref{sec:analysis-results}. 
The transitions are divided into two groups: the low-$J$ lines, spanning $J=12\text{--}11$ to $J=32\text{--}31$ (Fig.~\ref{fig-hc5n-low}), and the high-$J$ lines, ranging from $J=33\text{--}32$ to $J=43\text{--}42$ (Fig.~\ref{fig-hc5n-high}).

\clearpage
\section{LTE fit results}
\label{appendix:fitresults}
In this appendix, we present the results of the LTE fitting procedure for the molecules studied in this work.

\begin{table*}[h!]
\begin{center}
\caption{Results from the LTE fitting procedure.}
\label{table-fit}
\begin{tabular}{c l l l l r}
\hline
\hline
Molecule & Component & FWHM	& $\varv_{\rm LSR}$	&$T_{\rm ex}$	& $N$	\\
& & (km s$^{-1}$) & (km s$^{-1}$) & (K) & ($\times$10$^{14}$ cm$^{-2}$)\\
\hline
\multicolumn{6}{c}{HC$_{3}$N}\\
\hline
HC$_{3}$N\tablefootmark{a} & C1 & 23.52$\pm$0.17 & 67.79$\pm$0.09 & 16.53$\pm$0.15 & 6.54$\pm$0.07\\
 & C2 & 7.2$\pm$0.6  & 73.5\tablefootmark{b}  &14.5$\pm$1.7 &  9$\pm$3\\
H$^{13}$CCCN & C1 & 23.5\tablefootmark{b} &66.92$\pm$0.12	&11.4$\pm$0.1	&0.380$\pm$0.004	\\
& C2 & 7.2 &72.53$\pm$0.16	&14.5$\pm$0.6	&0.39$\pm$0.04	\\
HC$^{13}$CCN & C1 & 23.5	&67.32$\pm$0.17	&11.03$\pm$0.16	&0.372$\pm$0.007	\\
 & C2 &7.2	&71.9$\pm$0.2	&14.5$\pm$0.9	&0.41$\pm$0.05	\\
HCC$^{13}$CN & C1 & 23.5		&66.63$\pm$0.15	&11.25$\pm$0.15	&0.407$\pm$0.006	\\
 & C2 & 7.2			&72.2		&17.2$\pm$1.3	&0.27$\pm$0.04	\\
H$^{13}$C$^{13}$CCN & C1 & 23.5 &66.8 &13.8		&$<$0.01\tablefootmark{c}	\\
& C2 & 7.2	&72.1 &14.5 &$<$0.018\tablefootmark{d}	\\
& & & & 17.2 &$<$0.022\tablefootmark{d}		\\
H$^{13}$CC$^{13}$CN & C1 & 23.5	&66.8$\pm$1.3	&10.9$\pm$1.4	&0.0117$\pm$0.0014	\\
& C2 &  7.2	&72.1 &14.5 &$<$0.013\tablefootmark{e}		\\
& & & & 17.2 &$<$0.015\tablefootmark{e}		\\
HC$^{13}$C$^{13}$CN & C1 & 23.5 &66.8	& 17$\pm$3	&0.0095$\pm$0.0010\\
 & C2 & 7.2	&72.1 &14.5 &$<$0.029\tablefootmark{f}		\\
& & & & 17.2 &$<$0.032\tablefootmark{f}		\\
\hline
\multicolumn{6}{c}{HC$_{5}$N}\\
\hline
HC$_{5}$N low-$J$\tablefootmark{g} & C1 & 21.0$\pm$0.2	&65.70$\pm$0.14	&19.0$\pm$0.2	&1.247$\pm$0.018\\
 & C2 & 7.2	&73.5	&36$\pm$9	&1.7$\pm$0.3	\\
HC$_{5}$N high-$J$ & C1 & 21.0 &67.2$\pm$0.2	&22.9$\pm$0.5	&1.12$\pm$0.07	\\
 & C2 & 7.2		&73.5	&36	&0.34$\pm$0.05	\\
H$^{13}$CC$_{4}$N  & C1 & 21.0	&65.7	&16.2$\pm$1.6	&0.0288$\pm$0.0013	\\
& C2 & 7.2		&73.5 & 36 & $<$0.098\tablefootmark{h} \\
HC$^{13}$CC$_{3}$N  & C1 & 21.0	&65.87$\pm$0.18	&21.0$\pm$0.7	&0.0355$\pm$0.0005\\
& C2 & 7.2 &73.5 & 36 & $<$0.098\tablefootmark{i} \\
HC$_{2}^{13}$CC$_{2}$N  & C1 & 21.0 &65.7 &19 	&0.0302$\pm$0.0010\\ 
& C2 & 7.2 &73.5 & 36 & $<$0.075\tablefootmark{h}\\
HC$_{3}^{13}$CCN & C1 & 21.0 &66.0$\pm$0.4	&19 &0.0347$\pm$0.0010	\\
& C2  & 7.2 &73.5 & 36 & $<$0.124\tablefootmark{i}\\
HC$_{4}^{13}$CN & C1 & 21.0 &65.7 &17.6 $\pm$0.5	&0.0324$\pm$0.0010\\
& C2 & 7.2 &73.5 & 36 & $<$0.060\tablefootmark{h}\\
\hline
\end{tabular}
\end{center}
\tablefoot{\tablefoottext{a}{Taken from Table 2 of \citet{colzi2024}.} \tablefoottext{b}{Parameters without errors are fixed in the fitting procedure, as explained in Sect.~\ref{sec:analysis-results} and in \citet{colzi2024}.} 
\tablefoottext{c}{All the transitions are contaminated by other species. Upper limits calculated by hand simulating the maximum intensity that can reproduce the observed spectra (see Fig.~\ref{fig-h13c13ccn-fit}).}\tablefoottext{d}{Upper limit calculated as explained in Sect.~\ref{sec:analysis-results}, taking into account the rms of 3 mK at 105 GHz.}\tablefoottext{e}{Taking into account the rms of 0.5 mK at 44 GHz.}\tablefoottext{f}{Taking into account the rms of 0.6 mK at 36 GHz.}
\tablefoottext{g}{Only the low-$J$ transitions are used to calculate the isotopic ratios, since the high-$J$ transitions are not detected for the $^{13}$C isotopologues.}\tablefoottext{h}{Taking into account the rms of 0.7 mK at 34 GHz.}\tablefoottext{i}{Taking into account the rms of 0.7 mK at 31 GHz.}}
\end{table*}

\clearpage
\section{Main chemical pathways leading to HC$_{3}$N and HC$_{5}$N}
\label{app:formations}

In this Appendix we present the main chemical pathways leading to the formation of HC$_{3}$N and HC$_{5}$N in the models presented in Sect.~\ref{sec:astrochemistry}.

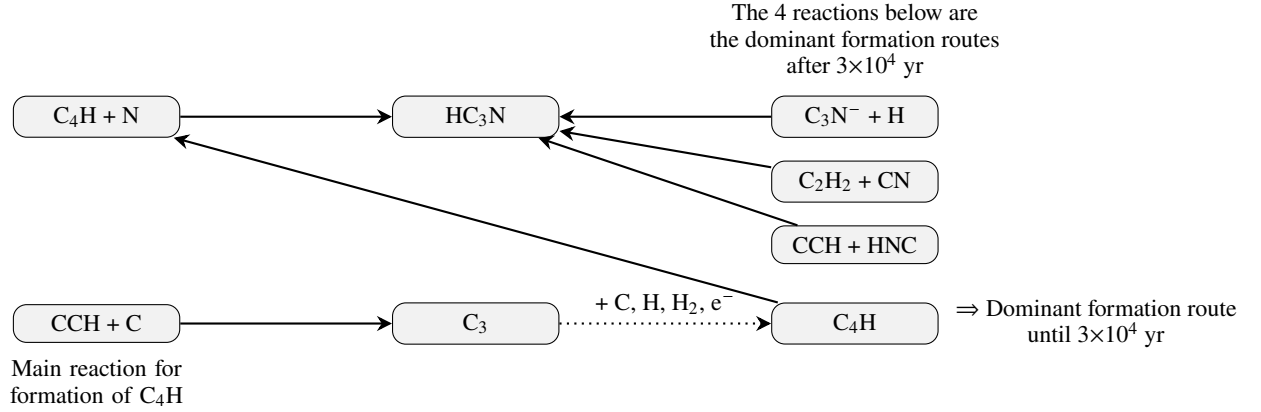
\begin{figure*}[h!]
\centering
\begin{tikzpicture}[
    node distance=2cm and 2.8cm,
    every node/.style={font=\small, align=center},
    species/.style={draw, rounded corners=5pt, fill=gray!10,
                    inner sep=4pt, minimum width=2.2cm},
    arrow/.style={-{Stealth[length=2mm,width=2mm]}, thick}
]

\node[species] (HC3N) {HC$_3$N};

\node[species, left=of HC3N] (C4HN) {C$_4$H + N};
\node[species, right=of HC3N] (C3NH) {C$_3$N$^-$ + H};
\node[species, below=0.3cm of C3NH] (C2H2CN) {C$_2$H$_2$ + CN};
\node[species, below=0.3cm of C2H2CN] (CCHHNC) {CCH + HNC};

\draw[arrow] (C4HN) -- (HC3N);
\draw[arrow] (C3NH) -- (HC3N);
\draw[arrow] (C2H2CN) -- (HC3N);
\draw[arrow] (CCHHNC) -- (HC3N);

\node[species, below=2.2cm of C4HN] (CCHC) {CCH + C};
\node[species, right=of CCHC] (C3) {C$_3$};
\node[species, right=of C3] (C4H) {C$_4$H};

\draw[arrow] (CCHC) -- (C3);
\draw[arrow,dotted] (C3) -- node[above]{+ C, H, H$_2$, e$^-$} (C4H);
\draw[arrow] (C4H) -- (C4HN);

\node[above=0.1cm of C3NH, font=\footnotesize]
    {The 4 reactions below are \\the dominant formation routes \\after 3$\times$10$^{4}$ yr};

\node[right=0.1cm of C4H, font=\footnotesize]
    {$\Rightarrow$ Dominant formation route \\until 3$\times$10$^{4}$ yr};

\node[below=0.1cm of CCHC, align=center, font=\footnotesize, text width=4cm]
    {Main reaction for formation of C$_4$H};

\end{tikzpicture}
\caption{Schematic representation of the main chemical pathways leading to the formation of HC$_3$N.
The dominant route proceeds via the neutral-neutral reaction C$_4$H + N, while additional
neutral and ion--neutral reactions contribute at later times. The dotted arrows indicate that this part of the pathway comprises multiple reactions that are not shown explicitly.}
\label{scheme-hc3n}
\end{figure*}

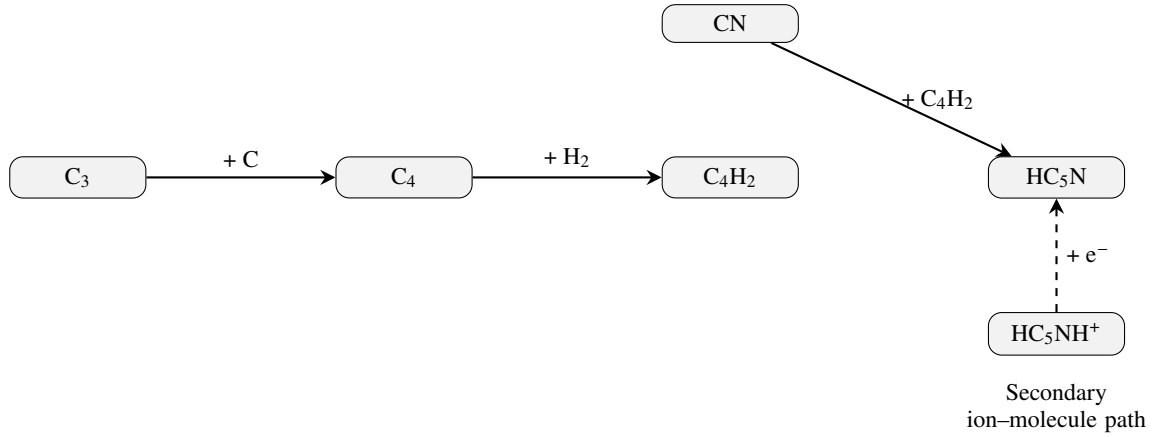
\begin{figure*}[h!]
\centering
\begin{tikzpicture}[
    node distance=2cm and 2.5cm,
    every node/.style={font=\small, align=center},
    species/.style={draw, rounded corners=5pt, fill=gray!10, inner sep=4pt, minimum width=1.8cm},
    arrow/.style={-{Stealth[length=2mm,width=2mm]}, thick}
]

\node[species] (C3) {C$_3$};
\node[species, right=of C3] (C4) {C$_4$};
\node[species, right=of C4] (C4H2) {C$_4$H$_2$};
\node[species, right=of C4H2] (HC5N) {HC$_5$N};

\node[species, above=1.5cm of C4H2] (CN) {CN};
\node[species, below=1.5cm of HC5N] (ion) {HC$_5$NH$^+$};

\draw[arrow] (C3) -- node[above]{+ C} (C4);
\draw[arrow] (C4) -- node[above]{+ H$_2$} (C4H2);
\draw[arrow] (CN) -- node[right]{+ C$_4$H$_2$} (HC5N);

\draw[arrow, dashed] (ion) -- node[right]{+ e$^-$} (HC5N);


\node[below=0.25cm of ion, align=center, font=\footnotesize, text width=3cm]
    {Secondary ion–molecule path};

\end{tikzpicture}
\caption{Simplified chemical network illustrating the main formation routes of HC$_5$N in the interstellar medium. 
C$_3$ contributes indirectly by forming longer carbon chains (C$_3$ → C$_4$ → C$_4$H$_2$), which then react with CN to produce HC$_5$N. 
The dashed arrow indicates an alternative ion–molecule pathway.}
\label{fig:hc5n_network}
\end{figure*}

\end{appendix}
\end{document}